\documentclass[pdflatex,sn-mathphys]{sn-jnl}

\usepackage[T1]{fontenc}
\usepackage[utf8]{inputenc}
\usepackage{lmodern}

\usepackage{color}
\usepackage{url}
\usepackage{natbib}
\usepackage{comment}
\usepackage{longtable}
\usepackage{soul}

\newcommand{\virg}[1]{``#1"}
\def \ergsc{\hbox{erg s$^{-1}$ cm$^{-2}$}}

\def \ergsch{\hbox{erg s$^{-1}$ cm$^{-2}$ Hz$^{-1}$}}

\def \nh {\hbox{ $N{\rm _H}$ }}
\def \colc {cm$^{-2}$}

\def \fwhm {{\em FWHM}}

\def \chis{$\chi ^{2}$}

\def \degr {\hbox{$^\circ$}}
\def\arcmin{\hbox{$^\prime$}}
\def\arcsec{\hbox{$^{\prime\prime}$}}
\def \rotdeg{$\degr/\rm day$}
\usepackage{mathtools}
\DeclarePairedDelimiter\abs{\lvert}{\rvert}
\newcommand{\sou }{Mrk~421}
\begin{document}


\title[Discovery of X-ray polarization angle rotation in active galaxy \sou]{Discovery of X-ray polarization angle rotation in active galaxy \sou}



\author[1]{Laura Di Gesu}
\author[2]{Herman L. Marshall}
\author[3]{Steven R. Ehlert}
\author[4,5,6]{Dawoon E.~Kim}
\author[1]{Immacolata Donnarumma}
\author[7]{Fabrizio Tavecchio}
\author[8]{Ioannis Liodakis}
\author[9,10]{Sebastian Kiehlmann}
\author[]{Iv\'{a}n Agudo \textsuperscript{11} }
\author[12,13]{Svetlana G. Jorstad}
% ...
\affil[1]{ASI \-Agenzia Spaziale Italiana, Via del Politecnico snc, 00133 Roma, Italy}
\affil[2]{MIT Kavli Institute for Astrophysics and Space Research, Massachusetts Institute of Technology, 77 Massachusetts Avenue, Cambridge, MA 02139, USA}
\affil[3]{NASA Marshall Space Flight Center, Huntsville, AL 35812, USA}
\affil[4]{INAF Istituto di Astrofisica e Planetologia Spaziali, Via del Fosso del Cavaliere 100, 00133 Roma, Italy}
\affil[5]{Dipartimento di Fisica, Universit\`{a} degli Studi di Roma \virg{La Sapienza}, Piazzale Aldo Moro 5, I\-00185 Roma, Italy}
\affil[6]{Dipartimento di Fisica, Universit\`a degli Studi di Roma \virg{Tor Vergata}, Via della Ricerca Scientifica 1, 00133 Roma, Italy}
\affil[7]{INAF Osservatorio Astronomico di Brera, Via E. Bianchi 46, 23807 Merate (LC), Italy}
\affil[8]{Finnish Centre for Astronomy with ESO, 20014 University of Turku, Finland}
\affil[9]{Institute of Astrophysics, Foundation for Research and Technology-Hellas, GR-71110 Heraklion, Greece}
\affil[10]{Department of Physics, University of Crete, 70013, Heraklion, Greece}
\affil[11]{Instituto de Astrof\'{i}sica de Andaluc\'{i}a-CSIC, Glorieta de la Astronom\'{i}a s/n, 18008, Granada, Spain}
\affil[12]{Institute for Astrophysical Research, Boston University, 725 Commonwealth Avenue, Boston, MA 02215, USA}
\affil[13]{Department of Astrophysics, St. Petersburg State University, Universitetsky pr. 28, Petrodvoretz, 198504 St. Petersburg, Russia}
% ...


\abstract{The magnetic field conditions in astrophysical relativistic jets can be probed by multiwavelength polarimetry, which has been recently extended to X-rays. For example, one can track how the magnetic field changes in the flow of the radiating particles by observing rotations of the electric vector position angle $\Psi$. Here we report the discovery of a $\Psi_{\mathrm x}$ rotation in the X-ray band in the blazar Mrk 421 at an average flux state. Across the 5 days of Imaging X-ray Polarimetry Explorer (IXPE) observations of 4-6 and 7-9 June 2022, $\Psi_{\mathrm x}$ rotated in total by $\geq360^\circ$. Over the two respective date ranges, we find constant, within uncertainties, rotation rates ($80 \pm 9$ and $91 \pm 8 ^\circ/\rm day$) and polarization degrees ($\Pi_{\mathrm x}=10\%\pm1\%$). Simulations of a random walk of the polarization vector indicate that it is unlikely that such rotation(s) are produced by a stochastic process. The X-ray emitting site does not completely overlap the radio/infrared/optical emission sites, as no similar rotation of $\Psi$ was observed in quasi-simultaneous data at longer wavelengths. We propose that the observed rotation was caused by a helical magnetic structure in the jet, illuminated in the X-rays by a localized shock propagating along this helix. The optically emitting region likely lies in a sheath surrounding an inner spine where the X-ray radiation is released.}

\maketitle


\section{Introduction} \label{sec:intro}

Despite decades of effort, the physical processes shaping the dynamics and emission of relativistic jets are in large part still unclear (e.g. \cite{blandford19}). However, in the last decade, several important clues have been obtained thanks to new facilities that allow the possibility of following in detail the time-variable emission properties of jets, especially at high photon energies. A similar leap is now expected from the newly-opened window of X-ray polarimetry, which provides us with an unprecedented view of the physical sites of particle acceleration and emission.
\\

Blazars, whose relativistic jets point close to the Earth, provide excellent laboratories to investigate the physics of jets and test current theoretical ideas (e.g. \cite{2017SSRv..207....5R}). Among blazars, the so-called high-peaked BL Lac objects (HBLs), whose synchrotron and Compton components peak in the X-ray and very-high-energy $\gamma$-ray band, respectively, are the ideal sources to investigate acceleration and cooling processes acting on ultra-relativistic electrons (with Lorentz factors $\gamma\gtrsim 10^6$). From the theoretical point of view, the main mechanisms advocated for the energization of particles in relativistic jets are diffusive shock acceleration (DSA; e.g., \cite{blandford87}), magnetic reconnection (e.g., \cite{sironi14}), possibly triggered by instabilities (e.g., \cite{bodo21}) or specific anti-parallel magnetic field structures \cite{zhang2020}, or reconnection plus stochastic acceleration in relativistic turbulence in a highly magnetized plasma (e.g., \cite{ComissoSironi2018}). 
\\
%
\begin{figure*}
\begin{minipage}[c]{1.0\textwidth}
 \includegraphics[width=1.0\textwidth]{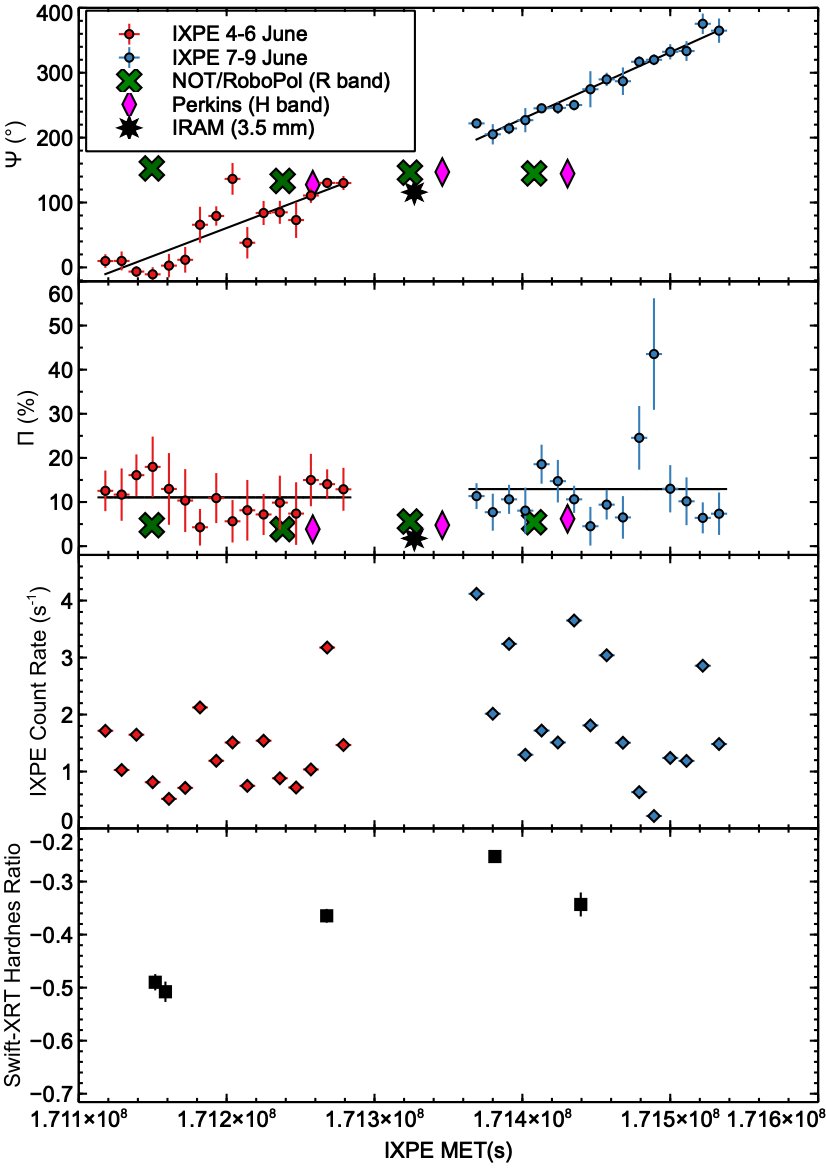}
\end{minipage}
\caption{From top to bottom: time evolution of the X-ray polarization angle, degree, IXPE photon count rate in time bins of $\sim$3 hours within the IXPE pointings of 4 June (red diamonds) and 7 June (blue diamonds) 2022, and hardness ratio in the X-ray band (black squares) measured with Swift-XRT. As a comparison, we display the polarization properties from simultaneous radio (IRAM: black stars), infrared (Perkins: magenta diamonds), and optical observations (NOT/RoboPol: green crosses).
}
\label{mvlcs.fig}
\end{figure*}

Polarimetry is a powerful tool that can break the degeneracies involved in the modeling of the spectral energy distribution (SED) and gain insight into the magnetic field and emission region geometry \cite{angel80}. Recent theoretical efforts have allowed us to identify polarimetric signatures expected under different scenarios of blazar emission regions that can be contrasted with observational evidence. Polarimetric measurements are very sensitive to the geometrical structure of the magnetic field permeating the flow and to its global properties (e.g., chaotic vs.\ globally ordered).  In particular, the evidence for systematic and large variations of the polarization angle, potentially associated with powerful $\gamma$-ray flares (e.g., \cite{blinov15,blinov18}), has been interpreted in terms of a helically twisted jet (e.g., \cite{larionov13}), an emission region moving along helical field lines that propagate down the jet (e.g., \cite{Vlahakis2006,marscher08,marscher10}), or light-travel delays when a shock forms in a jet with a predominantly helical/toroidal field  \cite{zhang2015,zhang16}. Such a structure for the global field is supported by (spatially resolved) polarimetric studies of parsec-scale jets in the radio band \cite{Hovatta2012,gabuzda21}. Interaction between a moving and a stationary shock can also produce large rotations \cite[e.g.,][]{Liodakis2020}. An alternative explanation is that the polarization behavior is stochastic, related to turbulence in the flow \cite{darcangelo2007,peirson18}, possibly combined with a standing shock \cite{marscher14,marscher15}. In this framework, the observed polarimetric parameters do not carry any direct information on the structure of the magnetic field in the jet, since they are mainly related to the turbulent nature of the flow. Turbulence can play an important role even if an average ordered magnetic field is present. In this case, while the polarization angle can display rotations related to the effect of the ordered field, the overall level of polarization can be decreased by the depolarizing effect of turbulence (e.g., \cite{marscher21,digesu2022}).
\\

\sou\ is a nearby (redshift $z=0.0308$) HBL that has been intensively studied at many wavelengths (e.g., \cite{abdo11}). It is among the first blazars detected at both GeV (by EGRET onboard the {\it Compton} Gamma Ray Observatory; \cite{lin1992}) and TeV energies (by the Whipple Observatory; \cite{punch1992}). It is bright and well-monitored in the X-ray band (e.g., \cite{Giommi2021,Middei2022}), where the synchrotron SED peaks at a high flux level, making it a prime target for linear polarization observations by IXPE.

%
%
\section{Results}\label{res.sec}
In the first year of IXPE operation, we observed the blazar \sou\ three times (on 4 May, 4 June, and 7 June 2022) to search for multi-epoch variability of the X-ray polarization properties. In this case, a single IXPE pointing extends over $\sim$2 days, which permits a search for intra-day variations of the polarization properties as well  \cite{digesu2022}. The IXPE observations were complemented with radio-millimetre [Very Long Baseline Array (VLBA), Institut de Radioastronomie Millimetrique (IRAM) and the SubMillimeter Array (SMA)], infrared (KANATA, Perkins Telescope), and optical polarization measurements [KANATA, Nordic Optical Telescope, Observatorio de Sierra Nevada (OSN), and Skinakas Observatory], as detailed in Table \ref{obs.tab} and displayed in Figures \ref{mvlcs.fig} and \ref{mv.fig}. We captured the source in an average activity state relative to its flux history \textbf{(see Methods)}, discounting occasional major, multi-band outbursts \cite{donna2009}.
\\

In May 2022, \sou\ was significantly X-ray polarized (with a polarization degree $\Pi_{\rm x} \sim 15\%$ at a position angle $\Psi_{\rm x} \sim 35 \degr$) and did not exhibit any substantial variability of the X-ray polarization properties within the IXPE pointing \cite{mrk421_ixped}. This is consistent with the X-ray polarization of Mrk~501, the other prototypical HBL observed by IXPE \cite{Liodakis2022-ixpe}. In both cases, the X-ray polarization degree was larger than at radio, infrared, and optical wavelengths. This suggested that the X-rays are produced by rapidly cooling, high-energy electrons accelerated at a shock, while the emission at longer wavelengths is emitted in a larger, downstream region containing a more disordered magnetic field and lower-energy electrons at increased distances from the shock.\\

In contrast, during two subsequent IXPE observations in June 2022, the X-ray polarization was undetected in the time-averaged IXPE data (see Methods section). In these observations, the X-ray flux was twice as high as in May 2022 ($\sim1.5 \times 10^{-10}$ and $\sim3.02 \times 10^{-10}$ \ergsc\ in the 2-8 keV IXPE band on 4-6 June and 7-9 June, respectively). A similar increase in flux (by a factor of $\sim$1.2; see Table \ref{obs.tab} and Fig. \ref{mv.fig}) was also measured in the radio (IRAM), infrared (Perkins),
and optical (NOT) data between 4-5 May and 6-7 June 2022.\\

 When measured over shorter time intervals of the IXPE observations of June 2022 (Fig.\ \ref{mvlcs.fig}), the X-ray polarization angle varied significantly in a manner consistent with a smooth rotation, as often observed for blazars in the optical band \cite[e.g.,][]{Marscher2008, Marscher2010,Blinov2015,Blinov2018,Blinov2021}. For example, on 4-6 June the X-ray polarization angle $\Psi_{\mathrm x}$ varied from $9\degr \pm 1\degr$ at the beginning of the IXPE pointing to $130\degr \pm11\degr$ at the end of the observation, while on June 7-9 $\Psi_{\mathrm x}$ changed from $222\degr \pm 7\degr$ to $360\degr \pm20\degr$ (considering time bins of $\sim$3 hours, as in Fig. \ref{mvlcs.fig}). We adopt a simple model where the polarization angle rotates at a constant rate, while the polarization degree remains constant, to test the hypothesis that a polarization angle rotation caused canceling of the polarization degree in the time-averaged data. Using a maximum-likelihood method, a fit of binned Stokes parameter time-series, and a fit in the Q-U plane (see Methods section), three independent analyses by subsets of the authors support this hypothesis. The data of 4-6 June are consistent with a rotation rate $\dot\Psi_{X}$ of $80 \pm 9 \degr/\rm day$ at constant $\Pi=10\%\pm1\%$, while for the data of 7-9 June the parameters are $\dot\Psi_{X}=91 \pm 8 \degr/\rm day$ and $\Pi_{\rm X}=10\%\pm1\%$. Our analysis suggests that both $\Pi_{\rm X}$ and $\dot\Psi_{X}$ probably varied somewhat with time about the above average values.\\

While the X-ray polarization angle was rotating, the millimetre-wave, infrared, and optical polarization angles did not vary substantially (Fig. \ref{mvlcs.fig}).  In the VLBA images of the radio core of 5 June (see Methods), some fanning out of the polarization vectors on the southeast side of the
the core is apparent, which indicates an azimuthal component to the magnetic field, as one would
expect if that component were helical.
In the optical, the scatter of $\Psi_{\mathrm O}$ during the June IXPE observations is $\sim20\degr$, with no significant trend. Given the gaps in the optical observations, and considering the intrinsic 180$\degr$ ambiguity, it is possible that a poorly sampled rapid rotation occurred. However, no straightforward scenario makes all the radio/infrared/optical data points consistent with the  time series of the X-ray polarization angle. Therefore, we conclude that polarization angle rotation at the longer wavelengths was either absent or proceeded at a different rate compared to the X-ray band.\\
Simultaneously with the polarization angle rotation, the X-ray spectrum changed significantly (see Fig. \ref{mvlcs.fig}, bottom panel, and Methods section). Swift X-ray Telescope (XRT) data taken over the same time frame as the IXPE pointing show that the flux in the 2.0-10.0 keV band increased by a factor of $\sim$3, while in the 0.3-2.0 band the increase in flux was by a factor of
$\sim$1.2. The maximum flux was recorded by the Swift-XRT in correspondence with the beginning of the second IXPE pointing. Then, by the end of the IXPE observation, the source brightness relaxed back to the same level as in the beginning. The spectral shape also changed in a harder-when-brighter fashion, with the brightest point corresponding to the flattest (hardest) spectrum. In contrast, in May 2022 the X-ray flux during the IXPE pointing changed in a similar fashion, but with no change in X-ray spectral shape or polarization angle \cite{mrk421_ixped}.
At $\sim$GeV $\gamma$-ray energies, \sou\ was in a quiescent state during the IXPE pointing: the Fermi Large Area Telescope (LAT) \citep{2009ApJ...697.1071A} observed minor activity in the 6-month period around
the IXPE observations, never reaching more than twice the average catalog flux (see fig. \ref{fig:latlc}).\\ 
\\
%

%
\begin{figure*}
\begin{minipage}[c]{1.0\textwidth}
\includegraphics[width=0.5\textwidth]{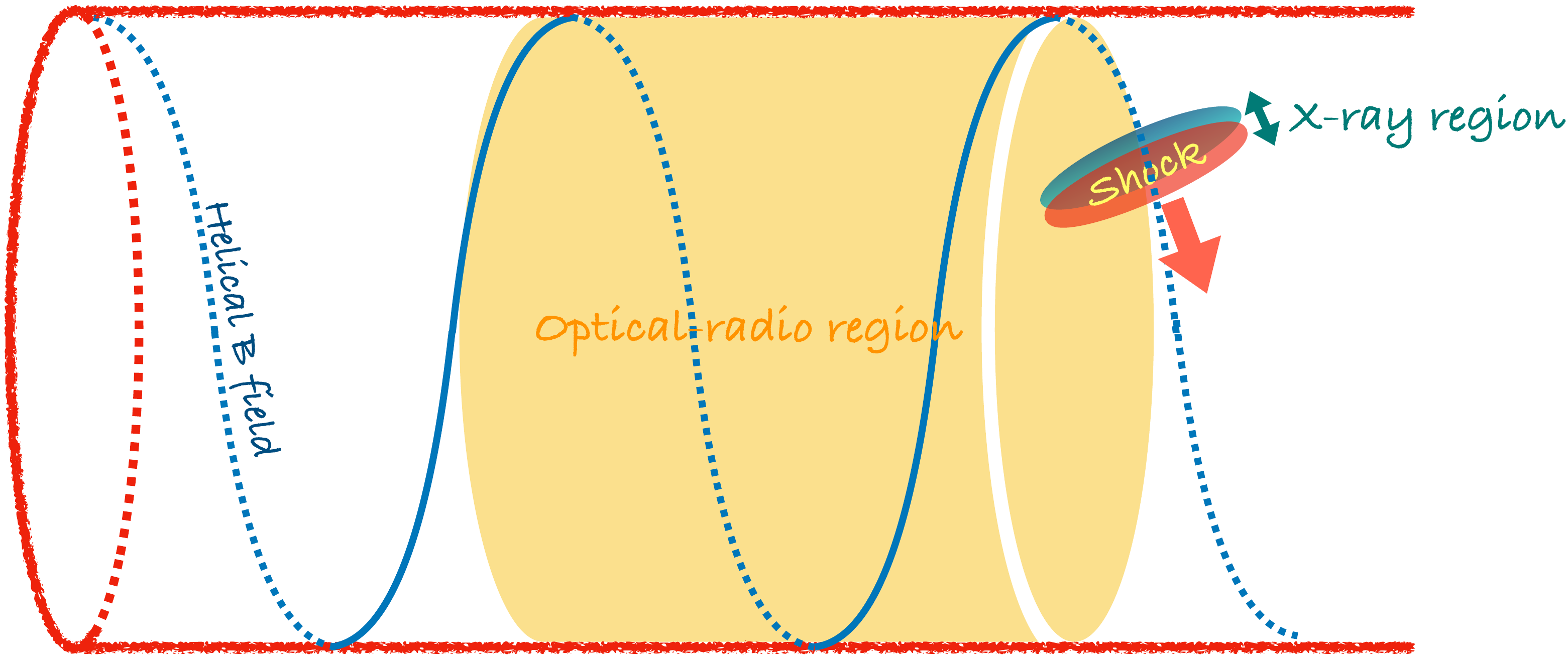}
 \includegraphics[width=0.4\textwidth]{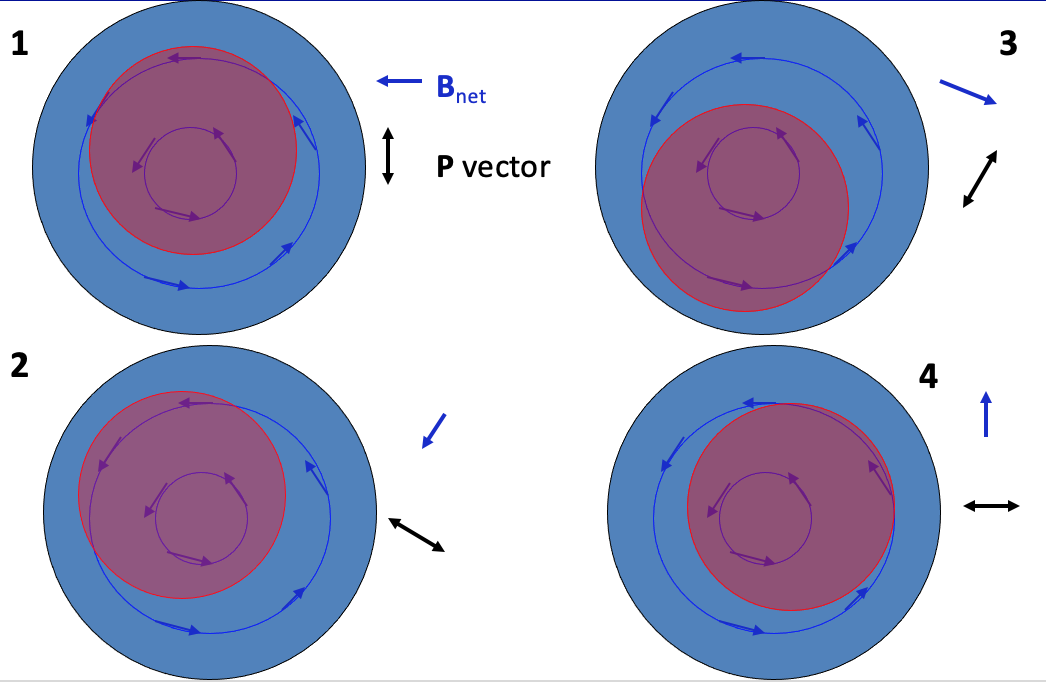}
\end{minipage}
\caption{Sketches of the scenario proposed to explain the X-ray polarization angle rotation in \sou.
Left: an off axis emission feature, e.g., a magnetosonic shock, propagates along helical magnetic field lines down the jet.
Right: the appearance of the emission feature, magnetic field, and polarization vector at 4 azimuthal positions along its spiral path as viewed by a distant observer aligned with the jet. The red circle represents the emission feature, while the blue-shaded region is the ambient jet.}
\label{cartoon.fig}
\end{figure*}

%
%
\section{Discussion}\label{sec:disc_conc}
Virtually all of the processes shaping the dynamics of particles in blazar jets predict specific characteristics of variability of the optical-to-X-ray polarization properties \cite{tavecchio21}, which we can compare against our observations of \sou. Our basic finding conforms to the scenario of an energy-stratified jet, as discussed above for \sou\ in May 2022 and for Mrk~501. We find that the roughly constant value of $\Pi_{\rm X}$, while the polarization angle was rotating, is higher than $\Pi_{\rm O}$, $\Pi_{\rm IR}$, and $\Pi_{\rm R}$, and similar to the level observed at non-rotating epochs. In addition, simultaneous radio, infrared, and optical polarization measurements do not show evidence of a similar polarization angle rotation.  Due to the limited sampling and intrinsic 180$\degr$ ambiguity, we cannot exclude the possibility that the optical polarization angle was rotating  at some level. However, we consider a rotation of the optical polarization angle at the same rate as the X-ray polarization angle unlikely. Indeed, typical optical polarization rotation rates of blazars vary between a few and a few tens of degrees per day \cite{kiehlmann2021}, in contrast
to the $\sim 85 \degr/\rm day$ that we estimated  in the X-ray. The optical rotation rates may be higher during bright outbursts associated with the passage of knots through the radio core \cite{magic2018,Liodakis2020}. No such outburst was recorded for \sou\ at the time of the IXPE pointings. Moreover, no solution of the 180$\degr$ ambiguity can completely reconcile the radio/optical/infrared polarization angle time series with the X-ray curve.
Therefore, as the polarization angle was likely not rotating (or rotating in a different way) at longer wavelengths, we suggest that the radio/infrared/optical emission sites are, at most, only partially co-spatial with the X-ray emission region.
\\

In an energy-stratified jet, the X-rays are emitted closer to the site of acceleration of the particles, while the lower-enegy particles emitting at radio to optical wavelengths span a larger region downstream owing to their longer radiative cooling times (e.g., \cite{Liodakis2022-ixpe}). Acceleration of electrons by a shock can explain the high X-ray polarization, since the shock front partially aligns a previously disordered magnetic field (e.g., \cite{hughes1985,tavecchio20}). The harder-when-brighter spectral behavior that \sou\ displayed while the X-ray polarization angle was rotating is also expected within the shock scenario \cite{kirk1998}.
\\

The energy stratification of the particles in the jet can be either linear, in the downstream direction, or radial, in the case of a structured jet \cite{georg2003,ghisellini2005,chhotray2017}, 
including an inner fast spine surrounded by a slower sheath. Such a spine-sheath jet explains, for example, the limb-brightening morphology that has been reported in very long-baseline interferometry (VLBI) radio maps of HBLs, including \sou\ \cite{giroletti2006, piner2010}.
In the first case, the turbulent component of the magnetic field could become increasingly dominant over the ordered component with distance from the shock \cite{marscher22}. In the second case, turbulence may prevail in the sheath. In both these frameworks, the particle-energy stratification would lead to lower polarization of the optical, infrared, and radio emission. Detailed modeling of the variability of flux and polarization induced by turbulence predicts erratic variability, as seen in the past at optical wavelengths in \sou\ \cite{marscher22}, which may by chance result in an apparent rotation of the polarization angle \cite{marscher14}. However, such apparent rotations are highly unlikely to explain, for example, the systematic rotations of the optical polarization angle in BL Lac and PKS~1510$-$089 \cite{Marscher2008, Marscher2010}. This has led to the proposal that these events occurred upstream of the most turbulent region of the jet. 
In a larger study of a number of optical-band rotation events in a sample of blazars, it was found that a random walk of the polarization vector cannot reproduce the majority of the observed properties of the population \cite{blinov15, Kiehlmann2017}.
\\

We have investigated whether the rotation of the polarization vector in \sou\ could be produced through random walks of the polarization angle by comparing the observations against simulated polarization light curves, following \cite{Kiehlmann2016,Kiehlmann2017} (see Methods). We consider two scenarios: either the rotations in  the observations of  4-6 June and 7-9 June are independent events, or they are part of a single long rotation that spanned seven days, of which we observed two segments separated by a one-day gap. In either scenario, we find, under the most favorable conditions, that only about 2\% of the simulated curves of $\dot\Psi_{X}$ and $\Pi_{X}$ vs.\ time are consistent with the observations. For the second scenario, if the rotation were actually longer than 7 days, the fraction of simulation trials that reproduce the observed properties would decrease. This suggests that it is unlikely that the observed behavior can be attributed to random variations of the polarization angle, and instead points to a coherent, deterministic process responsible for the observed rotation(s). 
\\

Magnetic reconnection episodes, which can result from the development of a large-scale kink instability \cite{Bodo2021} or current sheet \cite{zhang2021}, may also cause rapid polarization angle swings. These could be associated with $\gamma$-ray flares \cite{zhang2022} or quasi-periodic oscillations of the source flux \cite{dong2020,jorstad2022}. In our case, we did not witness any such oscillations or $\gamma$-ray flare while the polarization angle was rotating. Moreover, in contrast to our observations of \sou, simulations of magnetic reconnection often display, on average, a similar level of polarization degree in the X-ray and in the optical \cite{digesu2022}. In some cases, the X-ray polarization degree can also be much lower than the optical because of time averaging of the polarization vector, which is the opposite of what is observed \cite{digesu2022}.
\\

A specific model for deterministic polarization angle rotation involves an off-axis emission feature, such as a magnetosonic shock, propagating toward the observer and down a helical magnetic field \cite{marscher08,marscher10,larionov13,zhang16}. The observed rotation rate is determined by the time required for the feature to execute an orbit around the jet axis, modulated by light-travel delays and possibly other relativistic effects \cite{Peirson2018}. We display sketches of this  scenario in Fig. \ref{cartoon.fig}, both in the reference frame of the host galaxy (left panel) and as seen by an observer whose line of sight is along the jet axis (right panel). 
\\

As a quantitative example of such a model, we find that we can
reproduce the observed rotation rate $\dot\Psi_{X}\sim85$ \rotdeg\ over 5 days if (1) the center of
the emission feature is displaced by $2 -4\times10^{16}$ cm from the jet axis (for an angle to the line of sight ranging from $4^\circ$ to $0^\circ$), and (2) moves at a velocity of $0.99875c$ 
(Lorentz factor of 20), with a component parallel to the jet 
axis of $0.9975c$ and a transverse, rotational component of $0.05c$. During the rotation, the feature moves through a 
distance between 0.8 and 1.7 pc (depending on the line-of-sight 
angle) parallel to the line of sight in the observer's frame.
%
%
\\

If the jet is structured, the helical field component may be limited to the spine. Moreover, the spine can also contribute somewhat to the emission at longer wavelengths. This would lead to different polarization behaviour in X-ray than at longer wavelengths, and, at the same time, can explain a possible rotation of the optical polarization angle and the similarities in the variability of the X-ray to millimetre fluxes on longer time-scales (i.e., between May and June 2022).\\

The discovery of X-ray polarization angle
rotation in HBLs, made possible by the new capability of IXPE, opens up the possibility of fully investigating analogies and differences between polarization angle rotations at different frequencies and for different classes of blazars. For example, \cite{angelakis16} discuss a version of the model in which the polarization angle rotations occur immediately downstream of a shock, where the compressed magnetic field is dominated by an ordered (helical) magnetic field superposed on a turbulent component. As the shock is the site of acceleration of the particles that radiate at the frequency of the peak of the synchrotron SED, the highest polarization and main rotation events should be observed near that frequency. This argument
has been invoked to explain why optical polarization angle rotations preferably occur in low-synchrotron-peak (LSP) blazars \cite{blinov16}, and naturally predicts a more frequent occurrence of X-ray polarization angle rotations in the case of HBLs. In the same speculative framework, as optical polarization angle rotations are often correlated with $\sim$GeV $\gamma$-ray flares \cite{Blinov2018}, X-ray polarization angle rotations should be accompanied by $\sim$TeV activity, which is, in the case of HBLs, the analog of $\sim$GeV activity for LSPs.\\ 

In conclusion, this first study of X-ray polarization angle rotation in HBLs in a multi-wavelength framework shows how the availability of the X-ray polarization diagnostic enriches our ability to probe the magnetic field geometry and particle acceleration in different regions of relativistic jets, and thereby represents a new, fundamental step towards a comprehensive view of relativistic astrophysical jets.

\bmhead{Acknowledgments}

The Imaging X-ray Polarimetry Explorer (IXPE) is a joint US and Italian mission.  The US contribution is supported by the National Aeronautics and Space Administration (NASA) and led and managed by its Marshall Space Flight Center (MSFC), with industry partner Ball Aerospace (contract NNM15AA18C).  The Italian contribution is supported by the Italian Space Agency (Agenzia Spaziale Italiana, ASI) through contract ASI-OHBI-2017-12-I.0, agreements ASI-INAF-2017-12-H0 and ASI-INFN-2017.13-H0, and its Space Science Data Center (SSDC), and by the Istituto Nazionale di Astrofisica (INAF) and the Istituto Nazionale di Fisica Nucleare (INFN) in Italy.
This research used data products provided by the IXPE Team (MSFC, SSDC, INAF, and INFN) and distributed with additional software tools by the High-Energy Astrophysics Science Archive Research Center (HEASARC), at NASA Goddard Space Flight Center (GSFC).
  The IAA-CSIC group acknowledges financial support from the grant CEX2021-001131-S funded by MCIN/AEI/10.13039/501100011033 to the Instituto de Astrof\'isica de Andaluc\'ia-CSIC and through grant PID2019-107847RB-C44. The POLAMI observations were carried out at the IRAM 30m Telescope. IRAM is supported by INSU/CNRS (France), MPG (Germany), and IGN (Spain).
  The Submillimetre Array is a joint project between the Smithsonian Astrophysical Observatory and the Academia Sinica Institute of Astronomy and Astrophysics and is funded by the Smithsonian Institution and the Academia Sinica. Mauna Kea, the location of the SMA, is a culturally important site for the indigenous Hawaiian people; we are privileged to study the cosmos from its summit. Some of the data reported here are based on observations made with the Nordic Optical Telescope, owned in collaboration with the University of Turku and Aarhus University, and operated jointly by Aarhus University, the University of Turku, and the University of Oslo, representing Denmark, Finland, and Norway, the University of Iceland and Stockholm University at the Observatorio del Roque de los Muchachos, La Palma, Spain, of the Instituto de Astrofisica de Canarias. E. L. was supported by Academy of Finland projects 317636 and 320045. The data presented here were obtained [in part] with ALFOSC, which is provided by the Instituto de Astrofisica de Andalucia (IAA) under a joint agreement with the University of Copenhagen and NOT. We are grateful to Vittorio Braga, Matteo Monelli, and Manuel S\"{a}nchez Benavente for performing the observations at the Nordic Optical Telescope. Part of the French contributions is supported by the Scientific Research National Center (CNRS) and the French spatial agency (CNES). The research at Boston University was supported in part by National Science Foundation grant AST-2108622, NASA Fermi Guest Investigator grants 80NSSC21K1917 and 80NSSC22K1571, and NASA Swift Guest Investigator grant 80NSSC22K0537. This research was conducted in part using the Mimir instrument, jointly developed at Boston University and Lowell Observatory and supported by NASA, NSF, and the W.M. Keck Foundation. We thank D.\ Clemens for guidance in the analysis of the Mimir data. This work was supported by JST, the establishment of university fellowships towards the creation of science and technology innovation, Grant Number JPMJFS2129. This work was supported by Japan Society for the Promotion of Science (JSPS) KAKENHI Grant Numbers JP21H01137. This work was also partially supported by the Optical and Near-Infrared Astronomy Inter-University Cooperation Program from the Ministry of Education, Culture, Sports, Science and Technology (MEXT) of Japan. We are grateful to the observation and operating members of the Kanata Telescope. Some of the data are based on observations collected at the Observatorio de Sierra Nevada, owned and operated by the Instituto de Astrof\'{i}sica de Andaluc\'{i}a (IAA-CSIC). Further data are based on observations collected at the Centro Astron\'{o}mico Hispano en Andalucía (CAHA), operated jointly by Junta de Andaluc\'{i}a and Consejo Superior de Investigaciones Cient\'{i}ficas (IAA-CSIC). This research has made use of data from the RoboPol program, a collaboration between Caltech, the University of Crete, IA-FORTH, IUCAA, the MPIfR, and the Nicolaus Copernicus University, which was conducted at Skinakas Observatory in Crete, Greece. D.B., S.K., R.S., N. M., acknowledge support from the European Research Council (ERC) under the European Unions Horizon 2020 research and innovation program under grant agreement No.~771282. CC acknowledges support from the European Research Council (ERC) under the HORIZON ERC Grants 2021 program under grant agreement No. 101040021. 
  The research at Boston University was supported in part by National Science Foundation grant AST-2108622, NASA Fermi Guest Investigator grant 80NSSC21K1917 and 80NSSC22K1571, and NASA Swift Guest Investigator grant 80NSSC22K0537. This work was supported by NSF grant AST-2109127. We acknowledge the use of public data from the Swift data archive. Based on observations obtained with XMM-Newton, an ESA science mission with instruments and contributions directly funded by ESA Member States and NASA. Data from the Steward Observatory spectropolarimetric monitoring project were used. This program is supported by Fermi Guest Investigator grants NNX08AW56G, NNX09AU10G, NNX12AO93G, and NNX15AU81G. We acknowledge funding to support our NOT observations from the Finnish Centre for Astronomy with ESO (FINCA), University of Turku, Finland (Academy of Finland grant nr 306531). This work has made use of data from the Asteroid Terrestrial-impact Last Alert System (ATLAS) project. The Asteroid Terrestrial-impact Last Alert System (ATLAS) project is primarily funded to search for near earth asteroids through NASA grants NN12AR55G, 80NSSC18K0284, and 80NSSC18K1575; byproducts of the NEO search include images and catalogs from the survey area. This work was partially funded by Kepler/K2 grant J1944/80NSSC19K0112 and HST GO-15889, and STFC grants ST/T000198/1 and ST/S006109/1. The ATLAS science products have been made possible through the contributions of the University of Hawaii Institute for Astronomy, the Queen’s University Belfast, the Space Telescope Science Institute, the South African Astronomical Observatory, and The Millennium Institute of Astrophysics (MAS), Chile. The Very Long Baseline Array is an instrument of the National Radio Astronomy Observatory. The National Radio Astronomy Observatory is a facility of the National Science Foundation operated under cooperative agreement by Associated Universities, Inc.

\bmhead{Author contributions}


L. Di Gesu performed the analysis and led the writing of the paper. H. Marshall, S. R Ehlert, D. E. Kim, F. Muleri and A. Di Marco contributed to the IXPE data analysis. I. Donnarumma, F. Tavecchio, R.W. Romani A.\ P.\ Marscher, and S.\ G.\ Jorstad contributed to the discussion and other parts of the paper, and the latter two provided the VLBA images. I. Liodakis coordinated the multiwavelength observations.  I. Liodakis and S. Kiehlmann contributed the random walk simulations. I. Agudo, C. Casadio, J. Escudero, M. Gurwell, I. Myserlis, R. Rao and G. Keating contributed the radio observations. S.G. Jorstad contributed the infrared observations. S.S. Savchenko performed modeling of the host galaxy in H band.
B. Ag\'{i}s-Gonz\'{a}lez, F. J. Aceituno, I. Agudo, H. Akitaya, M. I. Bernardos, D. Blinov, G. Bonnoli, I. G. Bourbah, V. Casanova, Y. Fukazawa,  M. Garc\'{i}a-Comas, C. Husillos, R. Imazawa, K. S. Kawabata, S. Kielhmann, E. Kontopodis, P. M. Kouch, E. Lindfors, N. Mandarakas,  A. Marchini, T. Mizuno, T. Nakaoka, S. Romanopoulos, M. Sasada, R. Skalidis, A. Sota, M. Uemura, and A. Vervelaki,
contributed the optical observations. A.A. Vasilyev  has contributed to the paper discussion. R. Middei, M. Perri, and S. Puccetti contributed to the Swift data analysis. L. Pacciani and M. Negro contributed the Fermi data. The remaining authors are part of the IXPE team whose significant contribution made the X-ray polarization observations possible.

\bmhead{Competing interest}
The authors declare that they have no competing interests.

\bmhead{Data availability}
The data that support the findings of this study are freely available in the HEASARC IXPE Data Archive (\url{https://heasarc.gsfc.nasa.gov/docs/ixpe/archive/}).
The multiwavelength data are available on request from the individual observatories.

\def\aj{AJ}%
\def\actaa{Acta Astron.}%
\def\araa{ARA\&A}%
\def\apj{ApJ}%
\def\apjl{ApJ}%
\def\apjs{ApJS}%
\def\ao{Appl.~Opt.}%
\def\apss{Ap\&SS}%
\def\aap{A\&A}%
\def\aapr{A\&A~Rev.}%
\def\aaps{A\&AS}%
\def\azh{AZh}%
\def\baas{BAAS}%
\def\bac{Bull. astr. Inst. Czechosl.}%
\def\caa{Chinese Astron. Astrophys.}%
\def\cjaa{Chinese J. Astron. Astrophys.}%
\def\icarus{Icarus}%
\def\jcap{J. Cosmology Astropart. Phys.}%
\def\jrasc{JRASC}%
\def\mnras{MNRAS}%
\def\memras{MmRAS}%
\def\na{New A}%
\def\nar{New A Rev.}%
\def\pasa{PASA}%
\def\pra{Phys.~Rev.~A}%
\def\prb{Phys.~Rev.~B}%
\def\prc{Phys.~Rev.~C}%
\def\prd{Phys.~Rev.~D}%
\def\pre{Phys.~Rev.~E}%
\def\prl{Phys.~Rev.~Lett.}%
\def\pasp{PASP}%
\def\pasj{PASJ}%
\def\qjras{QJRAS}%
\def\rmxaa{Rev. Mexicana Astron. Astrofis.}%
\def\skytel{S\&T}%
\def\solphys{Sol.~Phys.}%
\def\sovast{Soviet~Ast.}%
\def\ssr{Space~Sci.~Rev.}%
\def\zap{ZAp}%
\def\nat{Nature}%
\def\iaucirc{IAU~Circ.}%
\def\aplett{Astrophys.~Lett.}%
\def\apspr{Astrophys.~Space~Phys.~Res.}%
\def\bain{Bull.~Astron.~Inst.~Netherlands}%
\def\fcp{Fund.~Cosmic~Phys.}%
\def\gca{Geochim.~Cosmochim.~Acta}%
\def\grl{Geophys.~Res.~Lett.}%
\def\jcp{J.~Chem.~Phys.}%
\def\jgr{J.~Geophys.~Res.}%
\def\jqsrt{J.~Quant.~Spec.~Radiat.~Transf.}%
\def\memsai{Mem.~Soc.~Astron.~Italiana}%
\def\nphysa{Nucl.~Phys.~A}%
\def\physrep{Phys.~Rep.}%
\def\physscr{Phys.~Scr}%
\def\planss{Planet.~Space~Sci.}%
\def\procspie{Proc.~SPIE}%
\let\astap=\aap
\let\apjlett=\apjl
\let\apjsupp=\apjs
\let\applopt=\ao




\section*{Methods}

\begin{table}[h]
\begin{center}
\begin{minipage}{400pt}
\caption{Log of polarization measurements of Mrk~421 used in this work.}\label{obs.tab}%
\begin{tabular}{@{}lllll@{}}
\toprule
Telescope & Band & Dates & Radio Flux Density & Radio Polarization\\
& (eV) & (YYYY-MM-DD) & ($10^{-25}$\ergsch{})  & $\Pi_{\rm R}$(\%) \quad $\Psi_{\rm R}$ ($\degr$)\\
\midrule
SMA & $9.5 \times 10^{-4}$ & 2022-06-04 & $28\pm 2$ & $  2.3\pm  0.2$ \quad $  130\pm    2$\\ 
 SMA & $9.5 \times 10^{-4}$ & 2022-06-16 & $38\pm 2$ & $  2.3\pm  0.2$ \quad $  133\pm   4$\\ 
 IRAM & $3.5 \times 10^{-4}$ & 2022-05-04 & $   28\pm    1$ & $  3.0\pm  0.7$ \quad $   237\pm    7$\\ 
 IRAM & $3.5 \times 10^{-4}$ & 2022-05-05 & $   38\pm    2$ & $  3.8\pm  0.7$ \quad $   232\pm    4$\\ 
 IRAM & $3.5 \times 10^{-4}$ & 2022-05-31 & $   32\pm    2$ & $  7.9\pm  0.8$ \quad $  128\pm    2$\\ 
 IRAM & $3.5 \times 10^{-4}$ & 2022-05-06 & $   36\pm    1$ & $  1.8\pm  0.6$ \quad $  116\pm    9$\\ 
 %
VLBA & $1.7 \times 10^{-4}$ & 2022-04-30 & $34.6\pm 0.1$ & $  1.6\pm  0.3$ \quad $100\pm 10$\\ 
 VLBA & $1.7 \times 10^{-4}$ & 2022-06-05 & $27.6\pm 0.1$ & $  2.6\pm  0.3$ \quad $147\pm 10$\\ 
 VLBA & $1.7 \times 10^{-4}$ & 2022-06-24 & $25.6\pm 0.1$ & $  2.2\pm  0.3$ \quad $132\pm 10$\\ 
 \midrule
Telescope & Band & Dates & Infrared flux density & Infrared Polarization \\
& (eV) & (YYYY-MM-DD) & ($10^{-25}$\ergsch{})  & $\Pi_{\rm IR}$(\%) \quad $\Psi_{\rm IR}$ ($\degr$)\\
\midrule
Perkins & H:0.9 & 2022-05-04 & $ 3.92\pm 0.05$ & $  1.3\pm  0.5$ \quad $   220\pm   11$\\ 
 Perkins & H:0.9 & 2022-05-05 & $ 1.77\pm 0.03$ & $  2.9\pm  0.8$ \quad $   212\pm    8$\\ 
 Perkins & H:0.9 & 2022-05-09 & $ 4.19\pm 0.05$ & $  3.1\pm  0.8$ \quad $    181\pm    7$\\ 
 Perkins & H:0.9 & 2022-05-14 & $ 3.55\pm 0.04$ & $  1.7\pm  1.0$ \quad $  175\pm   17$\\ 
 Perkins & H:0.9 & 2022-06-06 & $ 4.87\pm 0.05$ & $  3.9\pm  0.5$ \quad $  128\pm    3$\\ 
 Perkins & H:0.9 & 2022-06-07 & $ 4.52\pm 0.05$ & $  4.8\pm  0.9$ \quad $  147\pm    5$\\ 
 Perkins & H:0.9 & 2022-06-08 & $ 4.14\pm 0.06$ & $  6.2\pm  1.3$ \quad $  145\pm    6$\\ 
 Kanata \footnotemark[1] & H:0.9 & 2022-05-04 & $-$ & $  3.1\pm  0.2$ \quad $ 197\pm  2$\\ 
 Kanata \footnotemark[1] & H:0.9 & 2022-05-05 & $-$ & $  3.4\pm  0.3$ \quad $  187\pm  4$\\ 
 \midrule
 Telescope & Band & Dates & Optical flux density & Optical Polarization \\
& (eV) & (YYYY-MM-DD) & ($10^{-25}$\ergsch{})  & $\Pi_{\rm O}$(\%) \quad $\Psi_{\rm O}$ ($\degr$)\\
\midrule
 Skinakas & R:1.9 & 2022-05-13 & $ 1.60\pm 0.04$ & $  2.7\pm  0.2$ \quad $  192\pm    2$\\ 
 Skinakas & R:1.9 & 2022-05-13 & $ 1.64\pm 0.04$ & $  3.2\pm  0.2$ \quad $  195\pm    2$\\ 
 Skinakas & R:1.9 & 2022-05-13 & $ 1.61\pm 0.04$ & $  2.7\pm  0.2$ \quad $  193\pm    2$\\ 
 Skinakas & R:1.9 & 2022-05-14 & $ 1.48\pm 0.04$ & $  3.7\pm  0.2$ \quad $  179\pm    2$\\ 
 Skinakas & R:1.9 & 2022-05-16 & $ 1.41\pm 0.04$ & $  3.5\pm  0.3$ \quad $  176\pm    2$\\ 
 Skinakas & R:1.9 & 2022-05-18 & $ 1.41\pm 0.04$ & $  2.7\pm  0.4$ \quad $  179\pm    5$\\ 
 Skinakas & R:1.9 & 2022-05-22 & $ 1.53\pm 0.04$ & $  3.3\pm  0.2$ \quad $  118\pm    2$\\ 
 Skinakas & R:1.9 & 2022-05-30 & $ 2.02\pm 0.04$ & $  5.6\pm  0.2$ \quad $  149\pm    1$\\ 
 Skinakas & R:1.9 & 2022-05-31 & $ 1.98\pm 0.04$ & $  5.4\pm  0.2$ \quad $  150\pm    1$\\ 
 Skinakas & R:1.9 & 2022-06-01 & $ 2.05\pm 0.04$ & $  5.3\pm  0.2$ \quad $  142\pm    1$\\ 
 Skinakas & R:1.9 & 2022-06-02 & $ 1.93\pm 0.04$ & $  5.1\pm  0.2$ \quad $  152\pm    1$\\ 
 Skinakas & R:1.9 & 2022-06-07 & $ 2.12\pm 0.04$ & $  5.4\pm  0.3$ \quad $  145\pm    1$\\ 
 Skinakas & R:1.9 & 2022-06-17 & $ 2.05\pm 0.04$ & $  4.7\pm  0.3$ \quad $  152\pm    2$\\ 
 NOT & R:1.9 & 2022-05-04 & $ 1.89\pm 0.04$ & $  2.9\pm  0.1$ \quad $   202\pm    1$\\ 
 NOT & R:1.9 & 2022-06-04 & $ 2.02\pm 0.04$ & $  4.8\pm  0.1$ \quad $  153\pm    1$\\ 
 NOT & R:1.9 & 2022-06-05 & $ 2.25\pm 0.04$ & $  3.9\pm  0.1$ \quad $  133\pm    1$\\ 
 NOT & R:1.9 & 2022-06-06 & $ 2.27\pm 0.07$ & $  5.6\pm  0.2$ \quad $  146\pm    1$\\ 
SNO & R:1.9 & 2022-01-16 & $ 2.34\pm 0.01$ & $  3.8\pm  0.4$ \quad $  136\pm    3$\\ 
SNO & R:1.9 & 2022-04-19 & $ 1.95\pm 0.01$ & $  4.6\pm  0.7$ \quad $   264\pm    3$\\ 
SNO & R:1.9 & 2022-06-16 & $ 2.40\pm 0.50$ & $  9.5\pm  1.1$ \quad $  137\pm    4$\\ 
Kanata \footnotemark[1] & R:1.9 & 2022-05-04 & $ 2.50\pm 0.40$ & $  3.1\pm  0.2$ \quad $ 200\pm  1$\\ 
Kanata \footnotemark[1] & R:1.9 & 2022-05-05 & $ 2.20\pm 0.02$ & $  3.4\pm  0.3$ \quad $ 201\pm  1$\\ 
\midrule
Telescope & Energy range & Dates & X-ray flux & X-ray Polarization\footnotemark[2]\\
& (keV) & (YYYY-MM-DD) & ($10^{-11}$\ergsc)  & $\Pi_{\rm X}$(\%) \quad $\Psi_{\rm X}$ ($\degr$)\\
\midrule
 IXPE & 2.0-8.0 & 2022-05-04 & $ 8.67 \pm 0.03 $  & $15\pm2$ \quad $35\pm4$\\
 IXPE & 2.0-8.0 & 2022-06-04 & $ 15.69 \pm 0.09 $  & NOT DETECTED \\
 IXPE & 2.0-8.0 & 2022-06-07 & $ 30.2 \pm 0.2 $   & NOT DETECTED  \\
\botrule
\end{tabular}
\footnotetext[1]{Not corrected for dilution of polarization by unpolarized starlight from the host galaxy.}
\footnotetext[2]{X-ray polarization degrees are for the time-averaged IXPE data.}
\end{minipage}
\end{center}
\end{table}


\begin{table}[ht]
\begin{center}
\begin{minipage}{174pt}
\caption{Log of Swift observations of Mrk~421 in May/June 2022.}\label{xrtlog.tab}%
\begin{tabular}{@{}lll@{}}
\toprule
Swift-Obs ID & Dates & 0.3-10.0 keV flux \\
 & (YYYY-MM-DD) & ($10^{-11}$\ergsc)\\
\midrule
  31540031 & 2022-05-01 & $   46\pm    1$\\ 
   31540032 & 2022-05-03 & $   38\pm    1$\\ 
   31540033 & 2022-05-05 & $   44\pm    1$\\ 
   31540034 & 2022-05-07 & $   64\pm    1$\\ 
   96557003 & 2022-05-02 & $   46\pm    1$\\ 
   96557004 & 2022-05-03 & $   42\pm    1$\\ 
   96557005 & 2022-05-04 & $   39\pm    1$\\ 
   96557006 & 2022-05-05 & $   48\pm    1$\\ 
   96557008 & 2022-05-08 & $   84\pm    1$\\ 
   31540051 & 2022-06-01 & $   93\pm    1$\\ 
   31540052 & 2022-06-02 & $   90\pm    1$\\ 
   96557009 & 2022-06-03 & $   88\pm    1$\\ 
   31540053 & 2022-06-04 & $   75\pm    1$\\ 
   31540054 & 2022-06-04 & $   79\pm    1$\\ 
   31540055 & 2022-06-05 & $   69\pm    1$\\ 
   96557010 & 2022-06-06 & $  116\pm    1$\\ 
   89326001 & 2022-06-07 & $  141\pm    1$\\ 
   96557011 & 2022-06-08 & $  104\pm    2$\\ 
   96557012 & 2022-06-12 & $   73\pm    2$\\ 
   31540056 & 2022-06-16 & $   52\pm    1$\\ 
   31540057 & 2022-06-17 & $   69\pm    1$\\ 
   31540059 & 2022-06-19 & $  110\pm    1$\\ 
   31540060 & 2022-06-20 & $   74\pm    1$\\ 
   31540061 & 2022-06-21 & $   59\pm    1$\\ 
   31540062 & 2022-06-22 & $   61\pm    1$\\ 
   31540063 & 2022-06-23 & $   62\pm    1$\\ 
   31540064 & 2022-06-24 & $   56\pm    1$\\ 
\botrule
\end{tabular}
\end{minipage}
\end{center}
\end{table}

\subsection*{IXPE and multi-wavelength data} 
\label{sec:IXPE_data}

\subsubsection*{IXPE data}
The log of all of our observations of \sou\ in May-June 2022, including radio/millimetre, optical, infrared, and X-ray polarization measurements,  along with Swift monitoring  in the X-ray band, is listed in Table \ref{obs.tab} and \ref{xrtlog.tab}.\\
Here we report the analysis of the second and third IXPE observations of \sou\ 
that were performed from 11:20 UTC on 4 June
2022 to 11:00 UTC on 6 June (hereafter Obs.\ 2),
and from 09:00 UTC on 7 June to 11:10 UTC on 9 June (hereafter Obs. 3).
The livetimes for these pointings are $\sim$96\,ks and $\sim$86\,ks, respectively.\\

The raw IXPE data were processed using a standard pipeline\footnote{\url{https://heasarc.gsfc.nasa.gov/docs/ixpe/analysis/IXPE-SOC-DOC-009-UserGuide-Software.pdf}} that estimates the photo-electron emission direction, correcting for charging effects, detector temperature, Gas Electron Multiplier (GEM) gain non-uniformity, and instrumental spurious modulation \citep{rankin2022}. The level-2 event files (one for each of the three detector units (DU) on-board IXPE) from the pipeline contain all the information typical of an imaging X-ray astronomy mission (e.g., photon arrival time, detector coordinates, and energy), with the addition of the polarization information in the form of event-by-event Stokes parameters. Before proceeding with the scientific analysis, when necessary, we adjusted the data for small time-dependent changes to the gain correction obtained from data taken with the on-board calibration sources close to the actual time of observation \citep{Ferrazzoli2020JATIS}.\\

We performed the scientific analysis of the IXPE data using the ixpeobssim software \citep{pesce2019, baldini2022}. At the angular resolution of IXPE ($\sim30''$), blazars like Mrk~421 are point-like sources. 
Using the xpselect tool, we selected the source events using a circular 60\arcsec\ radius region, while we used an annulus centered on the source (with inner and outer radii of 2.5\arcmin\ and 5.0\arcmin) to estimate the background.  Within the ixpeobssim suite, the method of \cite{Kislat2015} for estimating the polarization of a set of events is implemented in the PCUBE algorithm.
We created the Stokes-parameter spectra of the source, using the PHA1, PHAQ, PHAU algorithms. The PHA1, PHAQ, and PHAU algorithms map the $I$, $Q$, and $U$ Stokes parameters of the photons into OGIP-compliant PHA spectral files (3 Stokes-parameter spectra per 3 detector units). We prepared the spectra for the spectro-polarimetric fit by binning the $I$ spectrum, requiring a minimum of 30 counts in each energy bin, as needed for the \chis\, statistics in the fits. Finally, we binned the $Q$ and $U$ spectra by grouping the channels by a constant factor of 10. 
To search for variability of the flux and polarization properties within the time span of the IXPE pointings, we created light curves with different time binning, using the LC algorithm in ixpeobssim. We applied the time binning of these light-curves to time-slice each of the event files
and computed a PCUBE in each time interval. We thereby created light curves of the polarization properties as a function of time, including normalized Stokes parameters ($q(t)$,$u(t)$), polarization degree ($\Pi(t)$), and polarization angle ($\Psi(t)$).
\\

We applied to the $\Psi(t)$ light-curves a standard procedure \citep{Blinov2015} to solve the 180\degr\ ambiguity of $\Psi$. By assuming that temporal variations of $\Psi$ are gradual,
we minimized the variation $\Delta_{\rm n} \Psi$ between consecutive points
$\Psi_{\rm n}$ and $\Psi_{\rm n+1}$ as follows.
For two consecutive data points $\Delta_{\rm n} \Psi$ was defined as 
$ \abs{\Psi_{\rm n+1}-\Psi_{\rm n}}-\sqrt{\sigma^2 (\Psi_{\rm n+1})+\sigma^2(\Psi_{\rm n})^2}$.
where $\sigma (\Psi_{\rm n+1})$ and $\sigma(\Psi_{\rm n})$ are the errors for the consecutive angles considered. If $\Delta_{\rm n} \Psi$
was larger than $90\degr$, we shifted $\Psi_{\rm n+1}$ by
multiples of $\pm180\degr$ as needed to minimize $\Delta_{\rm n} \Psi$. Otherwise, when $\Delta_{n}\leq90\degr$, we left $\Psi_{\rm n+1}$ unshifted. This procedure serves only for clarity of displaying the data, and is not critical for our estimation of the rotation rate, which was performed using unbinned event data or Stokes-parameter time series (see below).

\subsubsection*{$\gamma$-ray data}
\begin{figure*}
 \includegraphics[width=1.0\textwidth]{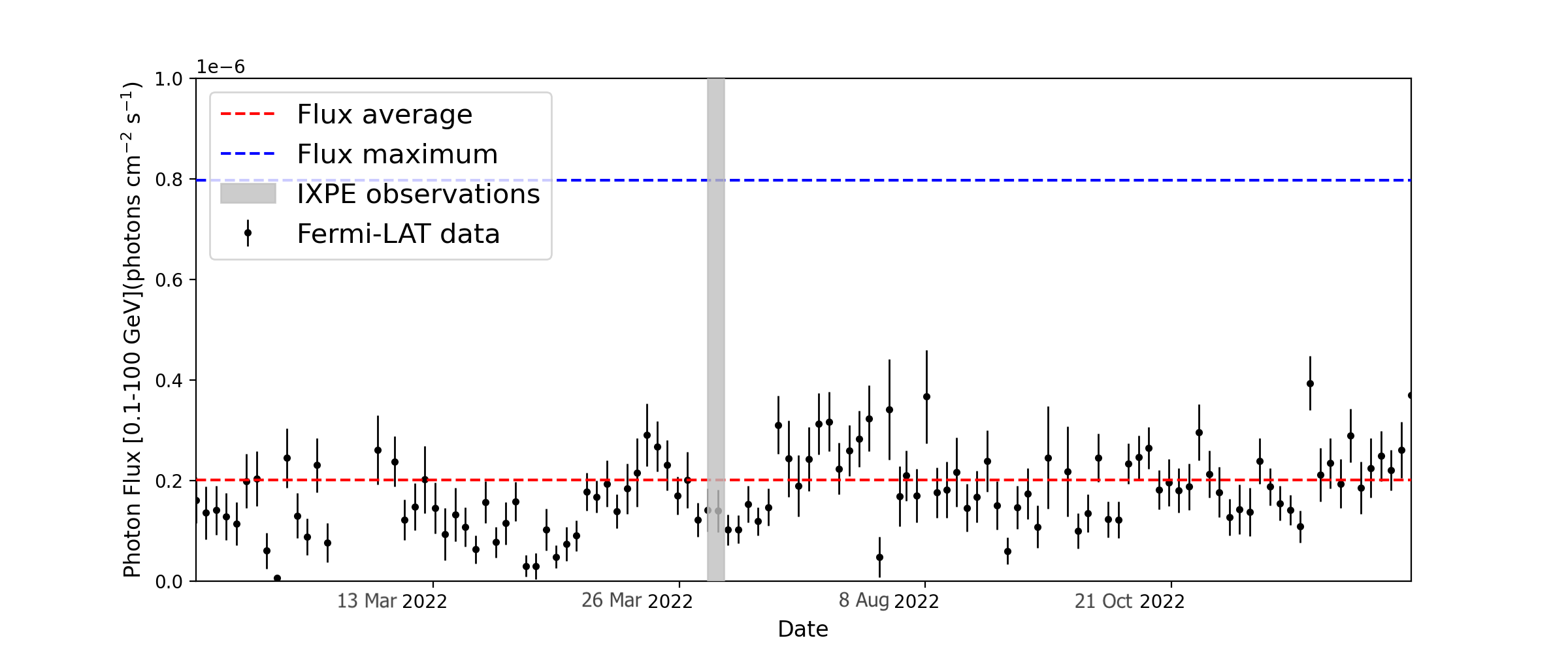}
\caption{Fermi-LAT light curve of \sou, integrated over the 0.1 -- 100 GeV energy range. The gray band illustrates the time range of the two IXPE observations of \sou\ discussed in this work. The red dashed line indicates the average photon flux of \sou\ over the LAT mission duration,
while the blue solid line marks the highest photon flux recorded by the LAT, which corresponds to the \sou\ flare of 2012. The plots include only points for which the test statistic for source detection is above 9.}
\label{fig:latlc}
\end{figure*}

We studied the $\gamma$-ray activity of \sou, analyzing data collected with the \textit{Fermi}-LAT  in the energy range 0.1-100 GeV.
We made use of the 3-day cadence light curve (within the years 2021 and 2022), which was extracted from the LAT Light Curve Repository (LCR), as displayed in Figure~\ref{fig:latlc}. We refer to \citep{LCR} for the detailed description of the unbinned maximum-likelihood data analysis. For this work, we adopted the light curve that was obtained using the option of leaving the spectral slope free to vary in the fit.
\\

\begin{figure}[ht]%
\centering
\includegraphics[width=0.9\textwidth]{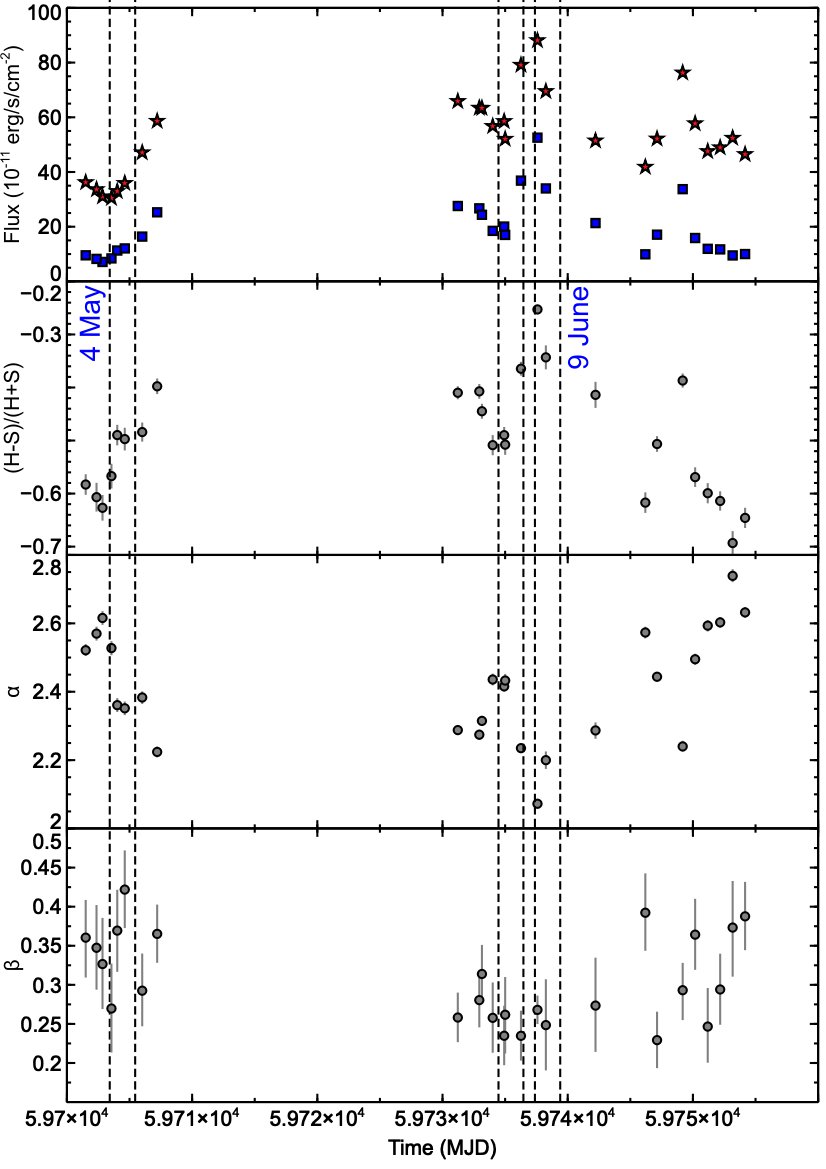}
\caption{Swift-XRT light curves of \sou\ in May-June 2022.
From top to bottom: soft (S band: 0.3-2.0 keV, red stars) and hard (H band: 2.0-10.0 keV, blue squares) fluxes, hardness ratio, $\alpha$ and $\beta$ parameters for a best-fit log-parabolic model. The dashed vertical lines indicate the beginning and the end of each IXPE observation.}\label{xrt.fig}
\end{figure}
\begin{figure}[ht]%
\centering
\includegraphics[width=0.9\textwidth]{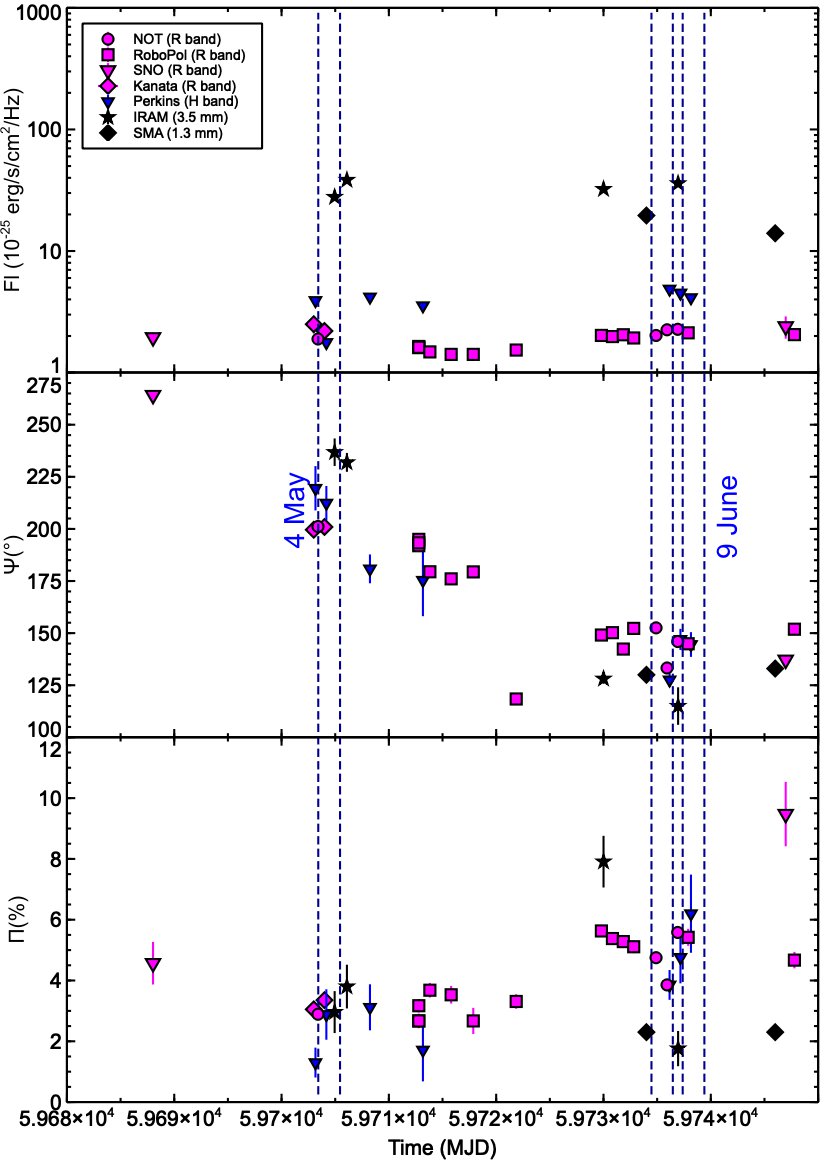}
\caption{Time evolution of the flux (top panel), polarization angle (middle panel), and polarization degree (bottom panel) of \sou\ at radio/millimetre (IRAM: black stars, SMA: black diamonds), infrared (Perkins: blue triangles), and optical (NOT: magenta circles, RoboPol: magenta squares, SNO: magenta triangles, Kanata: magenta diamonds) wavelengths. The dashed vertical lines mark the beginning and the end of each of the three IXPE observations.}
\label{mv.fig}
\end{figure}
\subsubsection*{X-ray flux and continuum spectrum}
We monitored the X-ray (0.3-10.0 keV) flux and spectral variations of \sou\ both near the time of, and simultaneously with, the IXPE pointings of May-June 2022 using the Neil Gehrels Swift X-Ray Telescope (XRT) (Table \ref{xrtlog.tab}).
The Swift-XRT observations each had an exposure time of $\sim1$ ks and were performed in Windowed Timing (WT) mode.
The data were processed using the XRT Data Analysis Software
(XRTDAS, v. 3.6.1). In the analysis, we used the latest calibration files
available in the {\it Swift-XRT} CALDB (version 20210915). The X-ray source spectrum was extracted from the cleaned event file using a circular region with a radius
of 47\arcsec. The background was extracted using a circular region with the same
radius from a blank sky WT observation available in the {\it Swift} archive. As a final step, we binned the 0.3-10 keV data to achieve at least 25 counts in each energy bin.\\

For our spectral analysis, we used the Xspec program \cite{xspec}. In all our fits we included Galactic absorption along the line of sight (\nh=$1.34 \times 10^{20}$ \colc; \cite{nh}) using the TBABS model, with Fe abundances set according to \cite{abund_wilm}.  For all the datasets, we found that a log-parabolic
model, where the photon index varies as a log-parabola in energy (i.e., $N(E)=K(E/E_{\rm p})^{(\alpha-\beta \log{(E/E_{p}}))}$ \cite{logpar}), fits the
spectra better than a simple power law (with $\Delta \chi^2$ in the range between
-23 and -233, for one additional degree of freedom). In the log-parabolic model, the pivot energy $E_{\rm p}$ is a scaling factor, $\alpha$ is the spectral slope at the pivot energy, $\beta$ corresponds to the spectral curvature, and $K$ is a normalization constant. In our fits, we set the pivot $E_{\rm p}$ to 5.0 keV \citep{balokovic2016,mrk421_ixped}, so that  $\alpha$ approximates the photon index in the 3.0-7.0 keV range. In this way, we extracted the X-ray spectral evolution of \sou\ in May/June 2022 (Fig. \ref{xrt.fig}), including soft (S band: 0.3-2.0 keV) and hard (H band: 2.0-10.0 keV) X-ray fluxes, hardness ratio (H-S/H+S), and $\alpha$ and $\beta$ parameters.
\\

Within the IXPE pointing of May 2022, the X-ray flux rose by a factor of $\sim$2 with no significant change in spectral shape \cite{mrk421_ixped}. Conversely, during the two IXPE pointings of June 2022, the flux increased from $\sim7.5 \times 10^{-10}$ \ergsc\ at the beginning of Obs. 2 to the brightest point of $\sim 1.4 \times 10^{-9}$ \ergsc\ at the beginning of Obs. 3, and then decreased back to the previous level. The maximum variation of the soft flux was by a factor of $\sim1.2$, while the hard flux varied by a factor of $\sim3$. This is in agreement with the IXPE measurement
in the 2.0-8.0 keV band, where the flux of Obs. 3 was $\sim$twice as high as during Obs. 2 (see Table \ref{obs.tab}). 
Therefore, the spectrum became harder and flatter (i.e., $\alpha$ decreased while the hardness ratio increased), with the flattest spectral shape coinciding with the brightest point.\\

\subsubsection*{Optical \& infrared data}
All the optical, infrared, and radio/millimetre polarization measurements of \sou\ in the time frame as the IXPE pointings
are displayed in Fig. \ref{mv.fig}.
Polarization observations in the optical were performed simultaneously with the IXPE pointings using the Hiroshima Optical and Near-InfraRed camera (HONIR) at the KANATA telescope, RoboPol at the Skinakas observatory 1.3-m telescope \citep{Ramaprakash2019}, the Alhambra Faint Object Spectrograph and Camera (ALFOSC) mounted at the 2.5m Nordic Optical Telescope (NOT), and the T90 telescope at the Sierra Nevada Observatory (SNO) during May/June 2022 in the R-band ($\lambda$=6500\AA/$E$=1.9 eV, \fwhm=1300 \AA /0.4 eV). The HONIR observations were performed in three optical bands\footnote{http://hasc.hiroshima-u.ac.jp/instruments/honir/filters-e.html}. The Stokes parameters are estimated by rotating the half-wave plate to four positions angles and combining the exposures. The instrument and data analysis are described in detail in \cite{Kawabata1999,Akitaya2014}. RoboPol is a novel 4-channel opto-polarimeter with no moving parts and with the ability to measure the normalized Stokes $q$ and $u$ parameters with a single exposure simultaneously. A detailed description of the semi-automated analysis and data reduction pipeline can be found in  \cite{Panopoulou2015,Blinov2021}.  The NOT data were reduced using the pipeline developed at the Tuorla Observatory, which follows standard photometric procedures \citep{Hovatta2016,Nilsson2018}. We performed the SNO observations with polarized $R_c$ filters, and calibrated the data using polarized and unpolarized standard stars. 
For the RoboPol, HONIR, and NOT observations in the R-band, we corrected for the host-galaxy depolarization by subtracting the (presumed unpolarized) host-galaxy flux density estimated for the individual apertures used in each observation \cite{Nilsson2007}): $\Pi_\mathrm{intr}= \Pi_\mathrm{obs}\times{I}/(I-I_\mathrm{host})$, where $\Pi_\mathrm{intr}$ and $\Pi_\mathrm{obs}$ are the intrinsic and observed polarization degree, respectively,  $I$ is the total flux density in mJy, and $I_\mathrm{host}$ is the host-galaxy flux density in mJy for the aperture of the observation \citep{Hovatta2016}.
\\

The IR observations were performed in H-band with the 1.8m Boston University Perkins telescope (Flagstaff, AZ) using the IR camera MIMIR in May/June 2022, supplemented by H-band HONIR observations in May 2022. The MIMIR instrument and data reduction procedures are described in detail in \cite{clemens2012}. The Perkins H-band observations were corrected for the host-galaxy contribution following the prescription above. To find the contribution of the host galaxy for the aperture necessary for the correction, we performed a photometric decomposition of the galaxy light using the IMFIT package \citep{Erwin2015}. The model for the image decomposition consisted of a Sersic profile for the host galaxy plus a point source for the blazar, determined by fitting a Moffat function \citep{Moffat1969} to several isolated stars in the image to create an average point-source function. We determined the parameters of the host galaxy as $n=1.18 \pm 0.35$ for the Sersic index, an ellipticity of $e=0.238 \pm 0.03$, an effective radius of $r_e=2.1'' \pm 0.14''$, and a surface brightness of $\mu_e = 16.59 \pm 0.17$. For the average aperture diameter of $5.2''$ used for the polarization observations, we estimated the contribution of the host-galaxy to be $12.41\pm0.06$ mag.
Any additional infrared observations have not been corrected for the host-galaxy contribution, and hence should be considered as lower limits to the intrinsic polarization degree. All the optical/infrared observations are listed in Table \ref{obs.tab}.
\begin{figure}[ht]%
\centering
\includegraphics[width=0.9\textwidth]{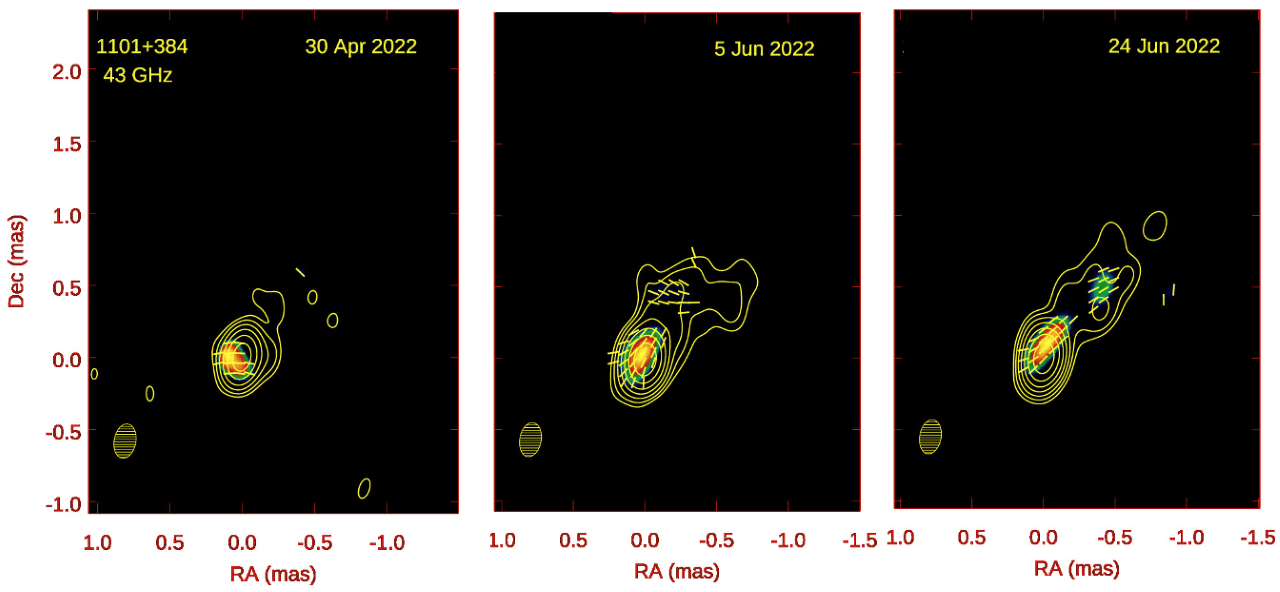}
\caption{VLBA images of \sou\ at 7 mm at three epochs near the times of the IXPE observations. Contours represent total
intensity, in factors of 2 starting at 2\% (left image) or 
1\% (middle and right images) of the peak intensity of (left to right) 0.35, 0.28, and 0.26 Jy per beam.
Colour coding indicates relative linearly polarized intensity, with
maxima of (left to right) 5.7, 7.2, and 5.7 mJy per beam. The yellow
line segments correspond to the position angle of polarization. The cross-hatched ellipse in the lower left of
each panel gives the FWHM of the elliptical restoring
beam, which corresponds to the direction-dependent angular
resolution. Marked positions are relative to the locations 
of the maximum intensity.}
\label{vlba.fig}
\end{figure}

\subsubsection*{Radio/millimetre data}
Images of \sou\ at 7 mm wavelength with a resolution of
$\sim0.2$ milliarcseconds (mas), corresponding to 0.12 pc, were produced with data from Very Long Baseline Array (VLBA) observations at epochs 30 April, 5 June, and 24 June 2022.  
The observations were from the
Blazars Entering the Astrophysical Multi-Messenger Era
(BEAM-ME) program (see website www.bu.edu/blazars/BEAM-ME.html). The data acquisition and analysis are described in
\cite{jorstad2017}. Figure \ref{vlba.fig} presents the images, in which total intensity is indicated by contours,
polarized intensity by colour coding, and polarization direction by yellow line segments. (Note that the 30 April
image has lower dynamic range than the others, hence the jet is not as apparent.) The images reveal a change in the
polarization direction of the ``core'' (feature of peak
intensity) between 30 April and the June epochs. The 5 June
image exhibits some fanning out of the polarization vectors
on the southeast side of the core, indicating an azimuthal
component to the magnetic field, as one would expect if that
component were helical. However, at the location of the
7 mm core, this component is weak relative to a more
randomly oriented magnetic field component.
\\

Polarization at the shorter wavelengths of 3.5 mm (86.24 GHz) and 1.3 mm (230~GHz) was measured with the 30~m Telescope of the Institut de Radioastronomie Millim\'etrique
(IRAM), located at the Pico Veleta Observatory (Sierra Nevada, Granada, Spain), on several nights in May/June 2022, as listed in Table \ref{obs.tab}. The observations were obtained within the Polarimetric Monitoring of AGN at Millimetre Wavelengths (POLAMI) program \citep{polami_i,polami_iii} following the reduction and calibration pipeline described in detail in \cite{polami_i}. 

No significant polarization at 1.3~mm from \sou\ was detected by IRAM either in May or June 2022, with an  upper limit (at a confidence level of 95\%) of $<6$\% and  $<4$\%, respectively. The early May 3.5 mm observations are consistent within uncertainties with an average $\Pi_R=3.4\pm0.7\%$ at an average $\psi_R=54.5\pm6^\circ$. The 30 May 2022 observation shows an increase in the polarization degree to $\Pi_R=7.9\pm0.9$, with a polarization angle almost perpendicular to the early May observations ($\psi=128\pm2.4^\circ$). The subsequent observation shows a similar $\Pi_R$ to the early May observations ($1.76\pm0.6\%$) at a roughly consistent angle of $\psi=116\pm9^\circ$.
We obtained additional observations at 1.3~mm (225.538~GHz) using the SubMillimeter Array (SMA) \cite{Ho2004} within the SMAPOL (\textbf{S}MA \textbf{M}onitoring of \textbf{A}GNs with \textbf{POL}arization) program. SMAPOL follows the polarization evolution of 40 $\gamma$-ray bright blazars, including Mrk\,421, on a bi-weekly cadence, as well as other sources in a target-of-opportunity (ToO) mode, including objects observed by IXPE. The \sou\ observations reported here were conducted on 4 June and 16 June 2022. The SMA observations employed two orthogonally polarized receivers, tuned to the same frequency range in full polarization mode, and used the SWARM correlator \cite{Primiani2016}. These receivers are inherently linearly polarized, but are converted to circular polarization by using the quarter-wave plates of the SMA polarimeter \cite{Marrone2008}. The lower sideband (LSB) and upper sideband (USB) covered 209-221 and 229-241 GHz, respectively. Each sideband was divided into six sub-bands with bandwidths of 2 GHz and a fixed channel width of 140 kHz. The SMA data were calibrated with the MIR software package \footnote{\href{https://lweb.cfa.harvard.edu/~cqi/mircook.html}{https://lweb.cfa.harvard.edu/~cqi/mircook.html}}. Instrumental polarization leakage was calibrated independently for USB and LSB using the MIRIAD task gpcal \cite{Saul1995} and removed from the data.  The polarized intensity, position angle, and percentage were derived from the Stokes I, Q, and U visibilities.
Between 4 and 16 June 2022, \sou\ decreased slightly in total flux, with values of 0.20$\pm$0.02 Jy and 0.14$\pm$0.02 Jy on 4 and 16 June, respectively. The linear polarization remained stable within the uncertainties between the two epochs, with a linear polarization degree of 2.3$\pm$0.2$\%$ and 2.3$\pm$0.3$\%$ and a polarization angle of 129.6$\pm$2.6$^{\circ}$ and 132.6$\pm$3.9$^{\circ}$
on 4 and 16 June, respectively. MWC\,349\, A was used for the total flux calibration, and the calibrator 3C\,286, which has a high linear polarization degree and stable polarization angle, was observed in both sessions as a cross-check of the polarization calibration.
%


\subsection*{X-ray Polarization Analysis}
\label{pol}
\subsubsection*{Time-averaged IXPE data}
\begin{figure*}[ht]%
\begin{minipage}[c]{1.0\textwidth}
 \includegraphics[width=0.5\textwidth]{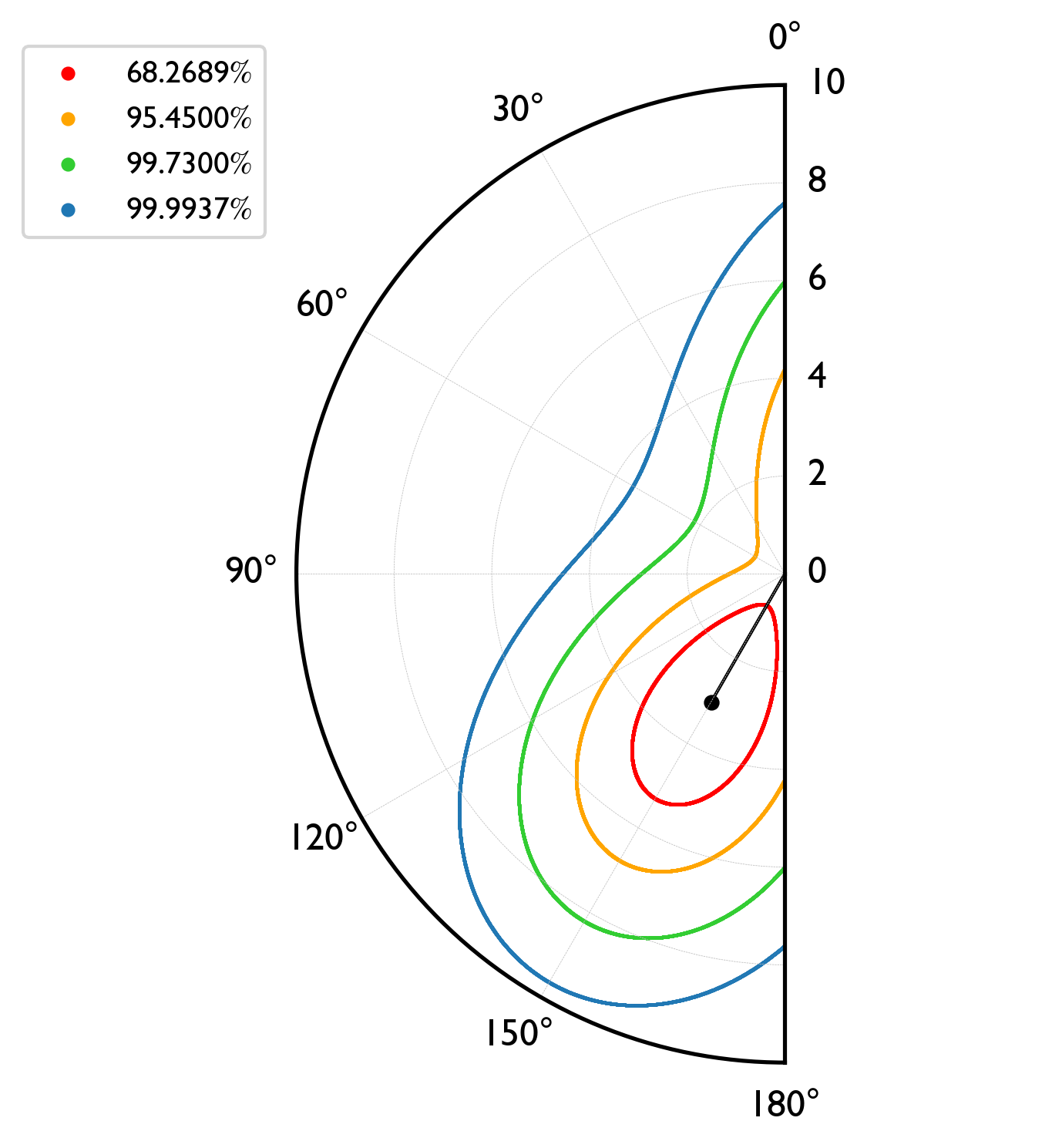}
 \includegraphics[width=0.5\textwidth]{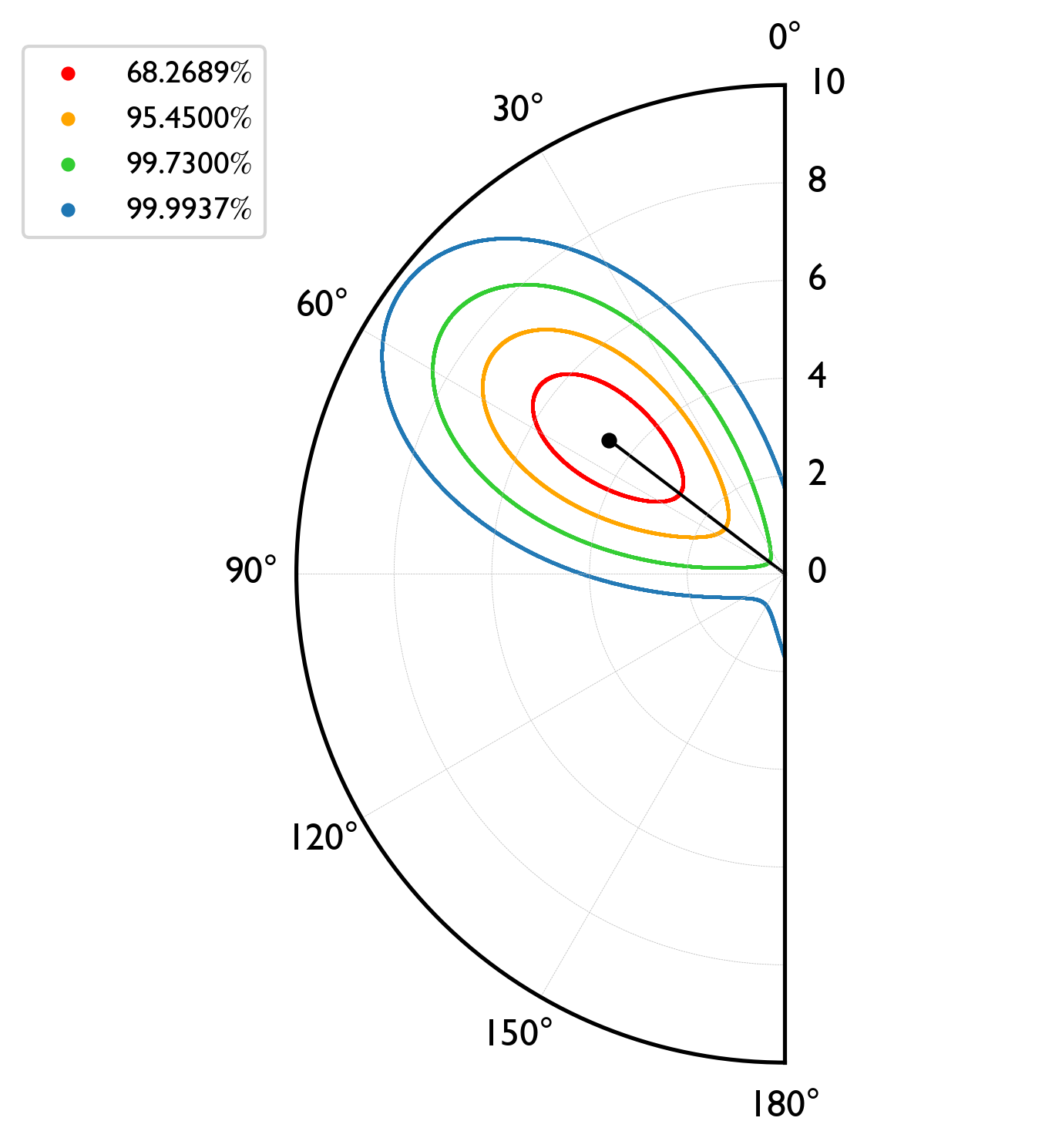}
\end{minipage}
 \caption{Polar plot of the $\Pi_{\rm x}$-$\psi_{\rm x}$ plane. Confidence contours of X-ray polarization derived within ixpeobssim for IXPE are shown for Obs.\ 2 (left panel) and 3 (right panel) of \sou.  The polarization degree is displayed in percentage units. For each case, the contours at 68.26\%, 95.45\%, 99.73\%, and 99.994\% confidence level (considering two degrees of freedom) are displayed, colour-coded in red, yellow, green, blue.}
  \label{cont.fig}
\end{figure*}

The analysis of the first IXPE observation of \sou, which discovered an X-ray polarization $\Pi_{X}=15\%\pm2\%$ at a polarization angle
$\Psi_{X}=35\degr \pm 4\degr$, is reported in
\cite{mrk421_ixped}. Conversely,
the X-ray polarization of \sou\ was undetected in the IXPE data of Obs.\ 2 and 3 when time-averaged over the entire length of the observations. Using the procedure of \cite{Kislat2015} within ixpeobssim, we found for the 2.0-8.0 keV band data a nominal minimum detectable polarization (equivalent to a 99\% confidence upper limit) of 5\% and 4\% for Obs.\ 2 and 3, respectively. Fig. \ref{cont.fig} displays the two-dimensional confidence contours for the X-ray polarization degree and angle obtained with the same method, to visually illustrate the lack of constraints on the X-ray polarization angle in these time-averaged data.
A spectropolarimetric fit \citep{Strohmayer2017} of the data gives similar upper bounds for the polarization properties, assuming that the polarization degree and angle are constant with energy (using the polconst model in Xspec). A log-parabolic model best fits the I spectra, where we used the same fixed pivot energy
as for the Swift-XRT data, $E_{\rm p}$=5.0 keV (see above). The IXPE data follow the same harder-when-brighter trend as for the Swift-XRT data: $\alpha=3.17\pm0.06$, $\beta=0.7\pm0.1$ and $\alpha=2.81\pm0.03$, $\beta=0.61\pm0.03$, for Obs.\ 2 and 3, respectively.\\

In the following, we investigate the cause of the change in the polarization properties of \sou.
The non-detection of the X-ray polarization of \sou\ in time-averaged IXPE
Obs.\ 2 and 3 data can be due to time variability of the polarization angle within the observation (see, e.g., \cite{digesu2022}). A rotation of the polarization angle, as often observed for blazars in the optical band, can in principle cancel the polarization degree in time-averaged data. In the following, we test this hypothesis using a simple model where the polarization angle rotates at a constant rate while the polarization degree remains constant. This assumption is sufficient to our aim, which is to assess the reality of the polarization angle rotation. Consideration of more complex models for the time variability is beyond the scope of this paper.
In order to make this assessment, we employ three methods: 
(1) a maximum-likelihood, unbinned event-based method, as in \cite{likelihood},
(2) a fit of the binned light-curves of the Stokes parameters using \chis\ statistics, and
(3) a fit of a rotation pattern in the Q-U plane using \chis\ statistics. 
A by-product of this analysis is an estimation of the rotation rate, for which all three methods return consistent results.

\begin{figure}[ht]%
\includegraphics[width=0.5\textwidth]{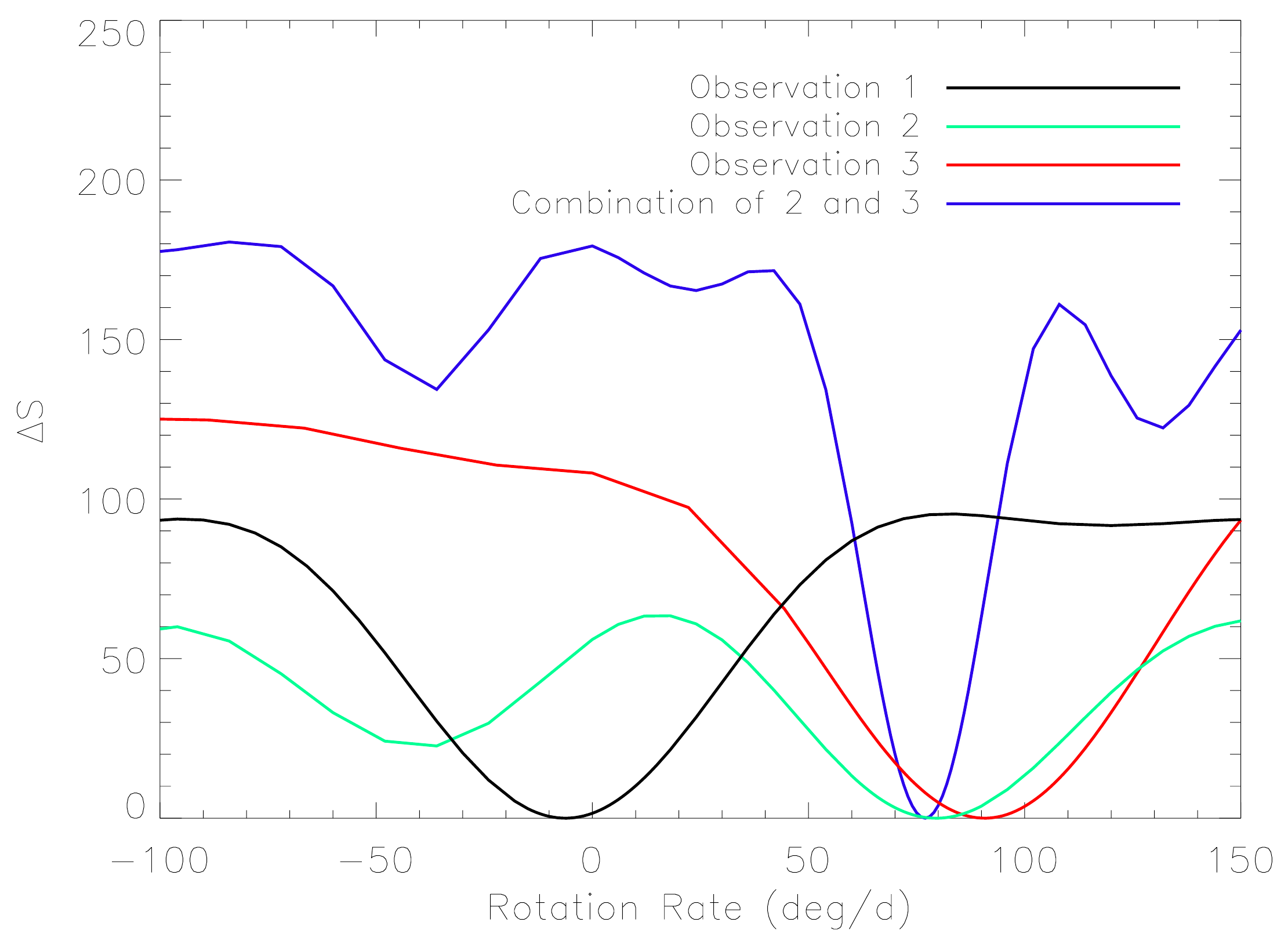}
 \caption{Log-likelihood difference as a function of the polarization angle rotation rate according to the maximum likelihood analysis method (see section \ref{sec:likelihood} and \cite{likelihood}).  For each observation set, the difference is relative to the minimum at the best fit value of the rotation rate, $\dot\Psi$.
 Results of the search for Obs. 1, 2, 3, and for
 Obs. 2 and 3 combined are colour-coded in black, green, red and blue, respectively. In Obs.\ 2, for example, $\dot\Psi=0$ can be ruled out at the 7 $\sigma$ level because $\Delta S > 50$ with respect to the best fit.}
  \label{likelihood.fig}
\end{figure}

%
\begin{figure*}
\begin{minipage}[c]{1.0\textwidth}
 \includegraphics[width=0.5\textwidth]{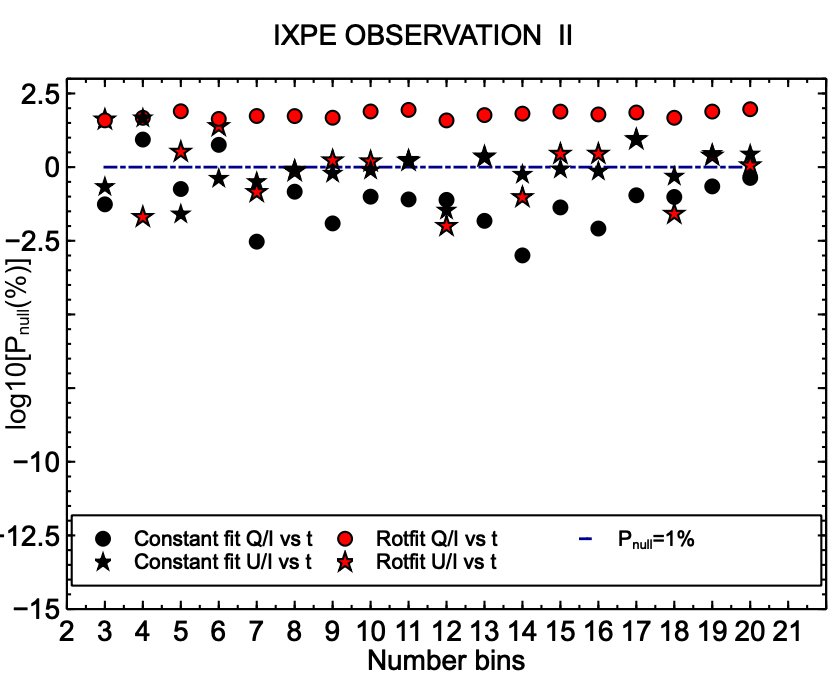}
 \includegraphics[width=0.5\textwidth]{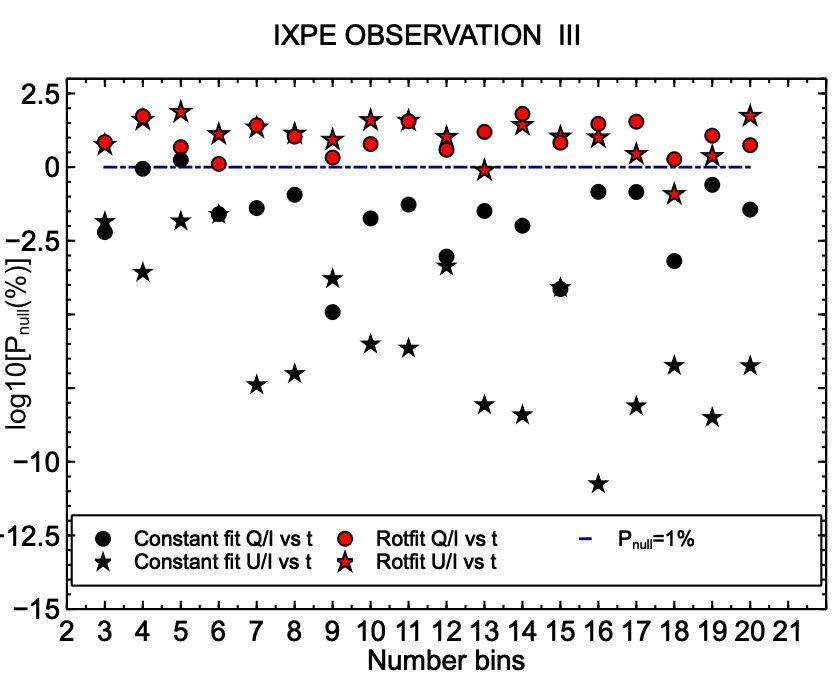}
\end{minipage}
 \caption{Results of fitting the Stokes parameters vs.\ time ($q(t)$: circles $u(t)$: stars), with a constant model (black symbols) or with a model including a constant rotation of the polarization angle (red symbols), for Obs.\ 2 (left panel)
 and 3 (right panel). The null hypothesis probability for all the fits is displayed as a function of the number of time bins involved.}
  \label{pnulls.fig}
\end{figure*}

\begin{figure*}[ht]%
\begin{minipage}[c]{1.0\textwidth}
\centering
\includegraphics[width=0.9\textwidth]{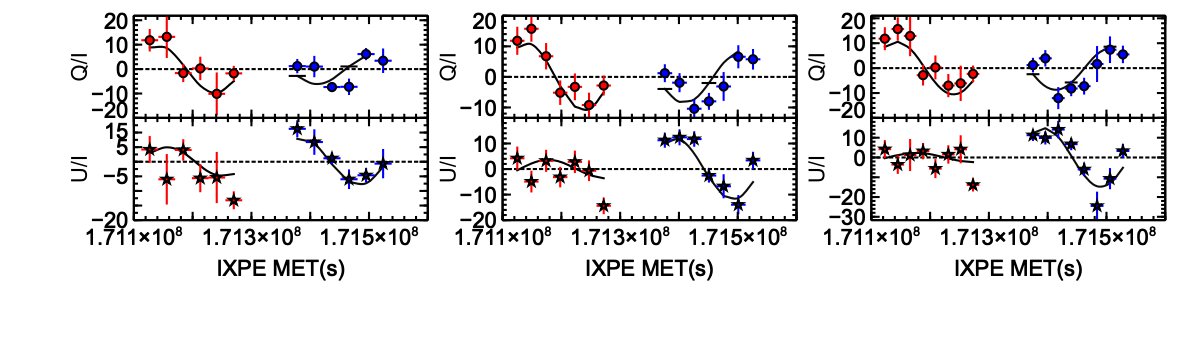}
\includegraphics[width=0.9\textwidth]{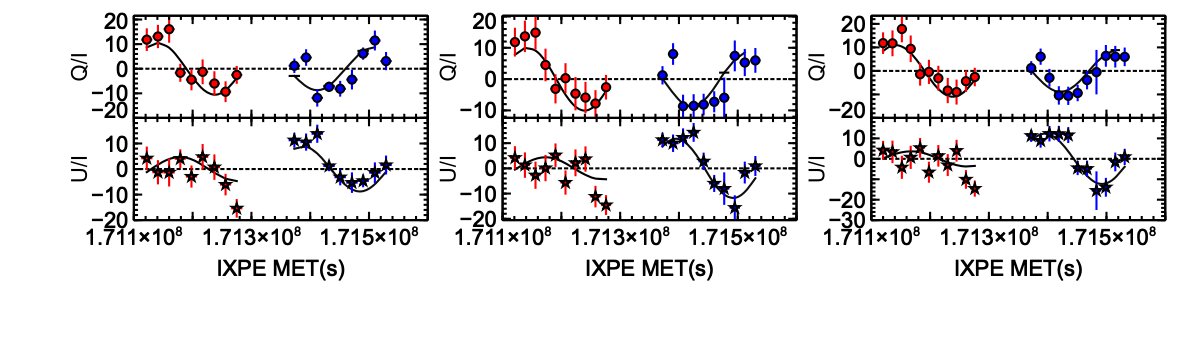}
\includegraphics[width=0.9\textwidth]{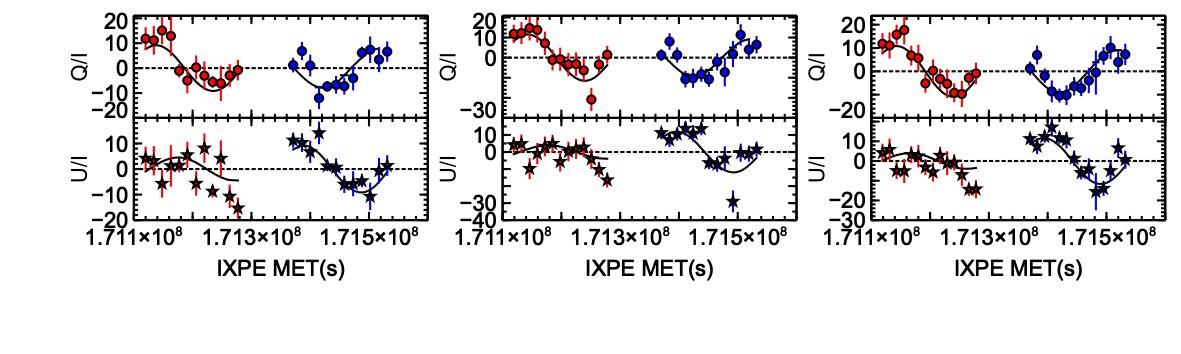}
\includegraphics[width=0.9\textwidth]{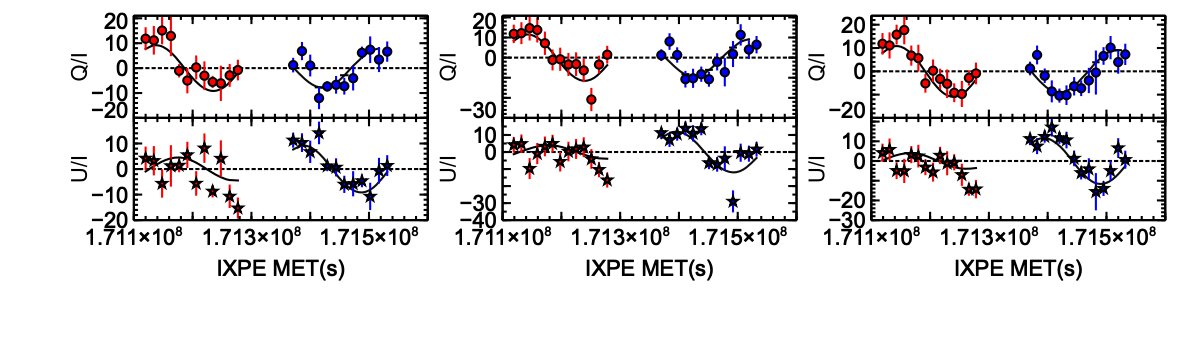}
\includegraphics[width=0.9\textwidth]{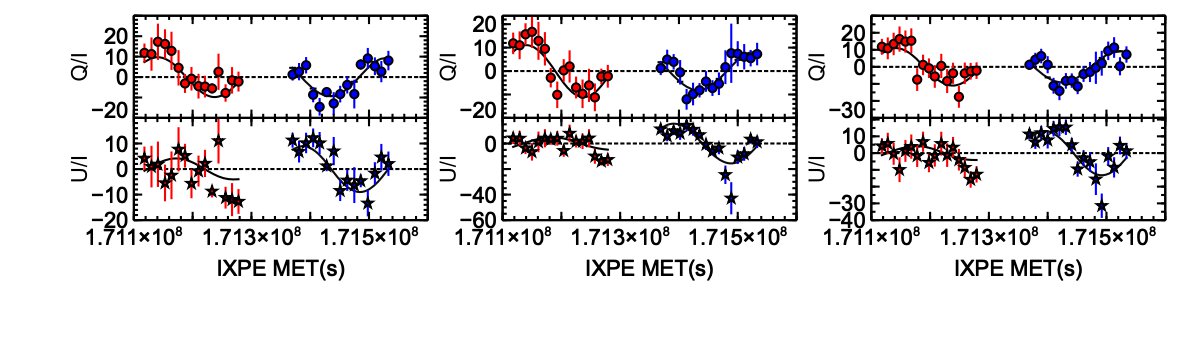}
\includegraphics[width=0.9\textwidth]{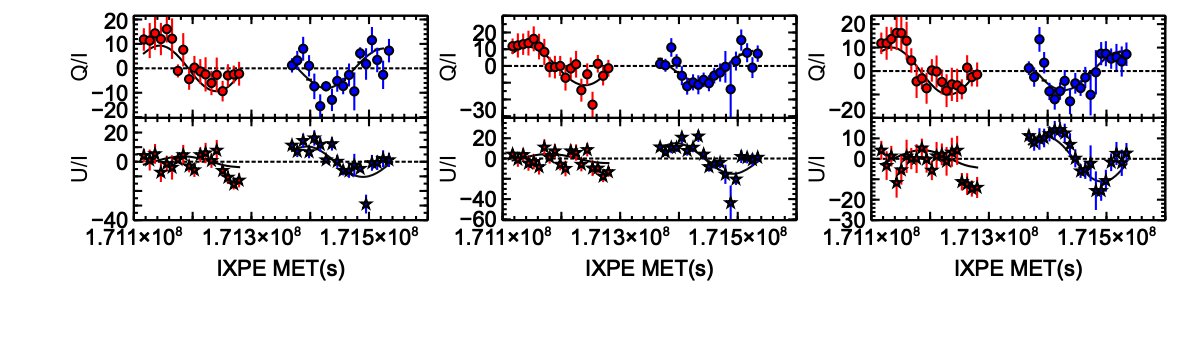}
\caption{Stokes parameters vs.\ time (($q(t)$: circles, top panel, $u(t)$: stars, middle panel) for Obs.\ 2 (red symbols) and 3 (blue symbols) for a number of time bins between 3 (i.e $\sim$ 17 hours) and 20 (i.e. $\sim$ one hour). The black solid line represents a fit with a model of a constant rotation of the polarization angle. The dotted horizontal line indicates null Stokes parameters.}
\label{stokesfit.fig}
\end{minipage}
\end{figure*}

\begin{figure}
    \centering
    \includegraphics[width=0.45\textwidth]{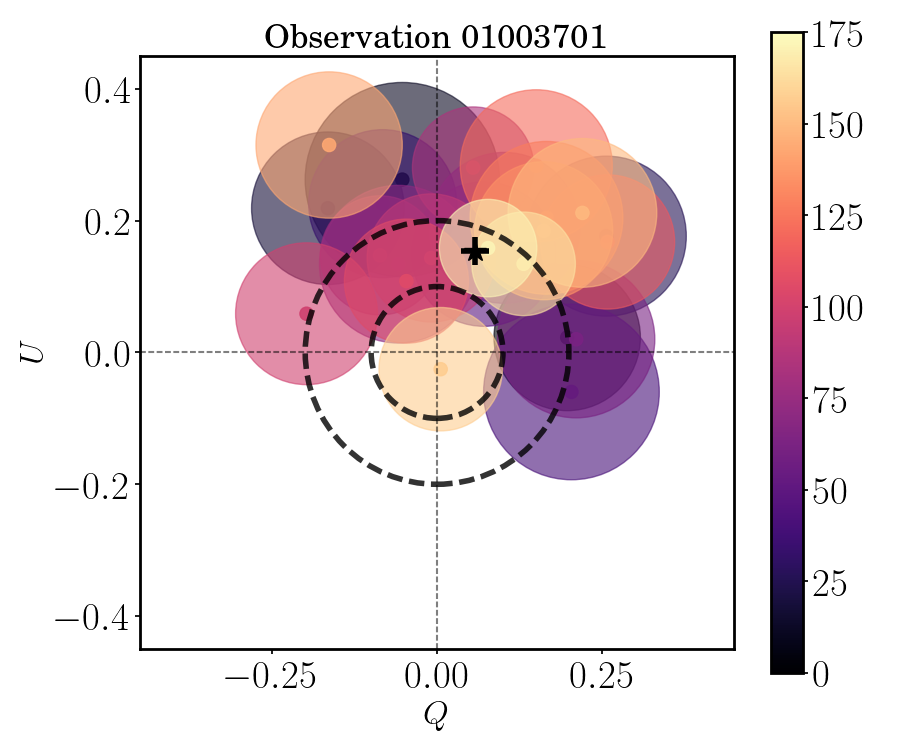}
    \includegraphics[width=0.45\textwidth]{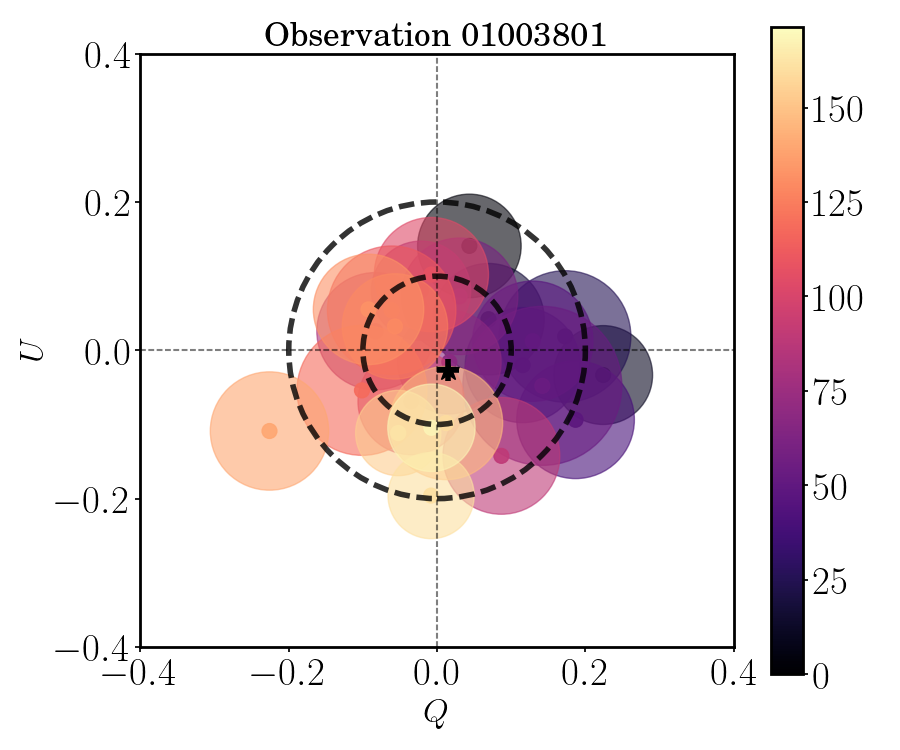}
    \includegraphics[width=0.45\textwidth]{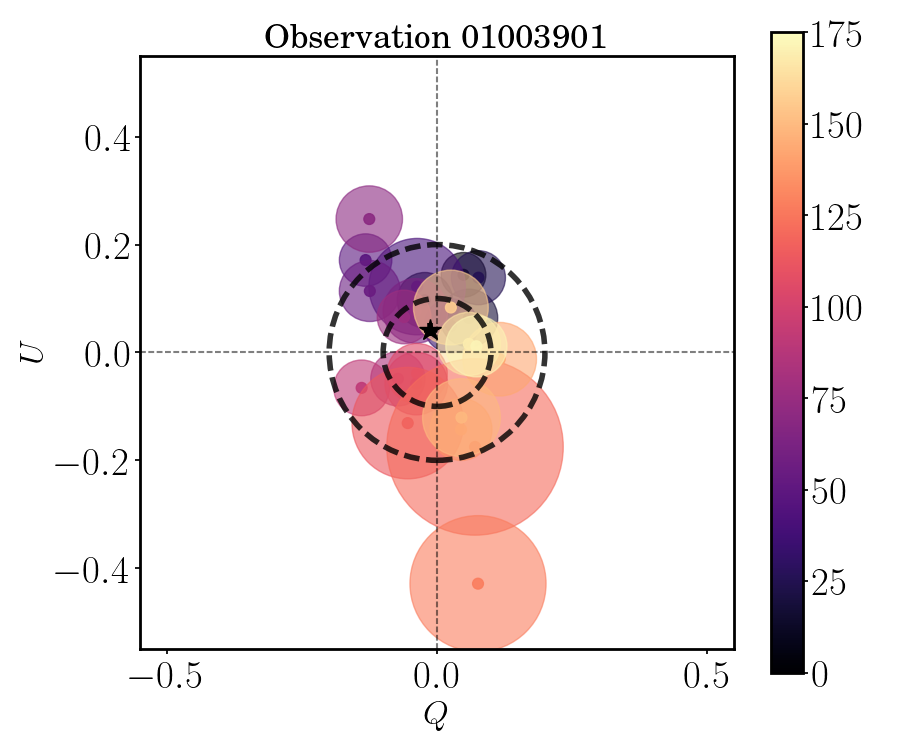}
    \caption{Variation in the Stokes parameters as a function of time through all three observations of \sou\ taken with IXPE. The black cross denotes the time-integrated average values of $Q$ and $U$ along with its $1-\sigma$ uncertainty region. The coloured circles denote the measured mean values of $Q$ and $U$ in each time bin and their $1-\sigma$ uncertainty radii.  The inner and outer dashed circles represent constant polarization degrees of $\Pi = 10\%$ and $\Pi = 20\%$, respectively, to help guide the eye. The colours of the data points correspond to the elapsed time relative to the beginning of the observation (in ks).  }
    \label{StokesEvolution.fig}
\end{figure}

\subsubsection*{Searching for the optimal rotation rate via a likelihood analysis}
\label{sec:likelihood}
For a detailed treatment of the optimal estimation of the (normalized) Stokes parameters $q$ and $u$ in the case of unbinned polarimetric data via
a likelihood analysis, we refer the reader to \citep{likelihood}. An unbinned likelihood analysis is appropriate for fitting X-ray event
data because of the Poisson nature of the counting statistics. This method has the advantage of being independent of any subjective choice of the binning of the data in energy or time.\\

Here, we expand the likelihood analysis method to the case of a rotating polarization angle. The log-likelihood formula for the polarization parameters of a set of (Poisson distributed) events depends on the Stokes parameters as
\begin{equation}
S(q,u)=-2\sum_{i} \ln(1+q\mu_{i} \cos 2\Psi_{i}+u\mu_{i} \sin2\Psi_{i}),
\end{equation}
where the sum is over each event $i$, $\mu_{i}$ is the modulation factor for event $i$ (at energy $E_{i}$), $\Psi_{i}$ is the event's instrumental phase angle, and it is assumed that $q$ and $u$ do not depend on the source flux. In addition, we are not fitting for the parameters of the spectral model.
\\

Since the time-averaged polarization is small and the apparent rotation is smooth, we can approximate the time behavior with a simple model that has
a varying polarization vector of constant degree, rotating at constant rate $\dot\Psi$. We also assume that there is no steady non-rotating component of the polarization (i.e., $q_0 = u_0 = 0$). For this model, the likelihood can be written as
\begin{equation}
S(\dot\Psi, q,u)=-2\sum_{i} \ln\{1+q\mu_{i} \cos [2 \dot\Psi(t_{i}-t_{0})]+u\mu_{i} \sin[2 \dot\Psi(t_{i}-t{_0})]\}, 
\label{eq:likelihood}
\end{equation}
where $t_{0}$ is a reference time, such as the midpoint of the observation period, and $t_{i}$ is the time of event $i$.
For any given rotation rate $\dot\Psi$, one can determine the optimal Stokes parameters 
$\hat{q}_r$ and $\hat{u}_r$ by marginalizing over $q$ and $u$ and substituting $\hat{q}_r$ and $\hat{u}_r$ back into Eq.~\ref{eq:likelihood} to yield $S'(\dot\Psi)$.
The best-fit value of $\dot\Psi$, $\hat{\dot\Psi}$, is given by the minimum of $S'(\dot\Psi)$ and, for Gaussian-distributed errors, uncertainties can be determined using $\Delta S'(\dot\Psi) = S'(\dot\Psi) - S'(\hat{\dot\Psi})$, since $\Delta S'(\dot\Psi)$ is distributed as
$\chi^2$ with one degree of freedom, being a function of only one interesting parameter.  Furthermore, we can compute the significance that $\hat{\dot\Psi}$ is non-zero using
\begin{equation}
\Delta S'(0) = S'(0) - S'(\hat{\dot\Psi}) = S(0, \hat{q}_{0}, \hat{u}_{0}) - S(\hat{\dot\Psi}; \hat{q}_{r}, \hat{u}_{r}).
\label{ds}
\end{equation}
Again, by marginalizing over $q$ and $u$ in order to test against the hypothesis that $\dot\Psi=0$, $\Delta S'(0)$ is distributed as $\chi^2$ with one degree of freedom.\\

We searched for the minima of $\Delta S(\dot\Psi)$ over a grid of values of $\dot\Psi$, which we set to be $-1000 \degr/\rm day$ to $+1000 \degr/\rm day$
using the IXPE event files of the three observations of \sou.
Upon determining the best-fit value of $\dot\Psi$, we estimated the uncertainty of the best-fit parameters at a confidence level of 90 \% using $\Delta S'(\dot\Psi)$.
The results of this analysis are displayed in Fig. \ref{likelihood.fig}. As expected, for Obs. 1, the estimated rotation rate is consistent with null rotation ($\dot\Psi=-6\pm9\degr/\rm day $), while the estimated polarization degree ($\Pi_{\rm x}=15\pm2 \%$) is consistent with our previous work \cite{mrk421_ixped}. Conversely, for Obs. 2 and 3 we find absolute minima of $\Delta S'(\dot\Psi)$ at $\dot\Psi=80\pm9\degr/\rm day$ and $\dot\Psi=91\pm8\degr/\rm day$, respectively, and that $\dot\Psi=0$ is ruled out at better than $7\sigma$ for both observations. The estimated polarization degrees for these two data sets were both $\Pi_{\rm x}=10\pm1 \%$, which suggests that while the polarization angle was rotating, the polarization degree may have remained similar to that measured when not rotating. The estimated rotation rates for Obs. 2 and 3 are consistent within the uncertainties, which suggests that we may have sampled two segments of one continuous rotation event spanning three days.  Combining Obs. 2 and 3, we found a rotation rate of $R=77.0\pm2.4\degr/\rm day$ and $\Pi_{\rm x}=10.1\pm0.8 \%$, indicating that the third observation was approximately in phase with the second after a day passed.
\subsubsection*{Modeling the time-binned Stokes parameters}
As a second test of the hypothesis of polarization angle rotation canceling the polarization degree in the time-averaged data, we modeled the time series of the normalized Stokes parameters. As in \cite{mrk421_ixped}, in order to investigate the polarization variability on different time scales, we used the temporal variations of the normalized Stokes parameters $q(t)$ and $u(t)$ with various time bin sizes in the range from $\sim$8 ks (corresponding to 
$N_{\rm bins}$=20 bins) to $\sim$60 ks ($N_{\rm bins}$=3).
We fit all the time series with (1) a constant model and (2) a model where the polarization angle rotates at a constant rate $\dot\Psi$ while the polarization degree $\Pi$ remains constant, such that the Stokes parameters vary as trigonometric functions: $q(t)=\Pi\cos(2(\dot\Psi t+\phi))$ and $u(t)=\Pi \sin(2(\dot\Psi t+\phi))$ (where $\phi$ is a phase angle). As a test of the goodness of the fits, we computed, for all the cases tested, the probability for the null hypothesis ($P_{\rm null}$) to obtain by chance a \chis\ value at least as large as that measured if the data are drawn according to the model.
\\

Figure \ref{pnulls.fig} displays, for both IXPE observations and for all the tested time binning schemes, the $P_{\rm null}$ values obtained for the constant model (black symbols) and the constant rotation model (red symbols). A constant model is statistically unacceptable ($P_{\rm null}<1\%$) for any time binning. A model including a linear time variation of the polarization angle always provides a better fit (i.e., an increase of $P_{\rm null}$, up to $\sim$99\% in several cases) than a constant model.
Figure \ref{stokesfit.fig} shows all the fit realizations, for the case $N_{\rm bins}$ in the range between 3 and 20. 
In the fitting exercise of Figure \ref{pnulls.fig}, we keep the parameters $\dot\Psi$ and $\Pi$ fixed to those determined from the likelihood analysis. Hence, $\phi$ is the only free parameter.
Moreover, we note that the independent fits of $q(t)$ and $u(t)$ return best-fit values of $\phi$ consistent within each other (within a typical uncertainty of $\sim8\degr$). This is an \textit{a posteriori} indication that our fits are physical. To further assess the phenomenology of the polarization variability, we allow, in turn, the rotation rate $\dot\Psi$ and the polarization degree $\Pi$ to be free in our fits. We find that $\dot\Psi$ is unconstrained in these fits. Conversely, the best fit corresponds to values of $\Pi$ that are marginally inconsistent between the fit of $q(t)$ (e.g., $\Pi=11\%\pm2\%$ for the case $N_{\rm bins}$=16) and that of $u(t)$ (e.g., $\Pi=5\%\pm2\%$). This suggests that $\Pi$ and/or $\dot\Psi$ are somewhat variable with time about the average values, rather than constant. The detailed modeling of these variability features is beyond the scope of our hypothesis testing, and may be assessed in future works in the context of physically motivated models.\\
\subsection*{Fitting a rotation model in the $Q$-$U$ plane}
As a third and final test, we modeled the time behaviour of the data in the $Q$-$U$ plane, with $N_{\rm bins}=24$. We display the time evolution of the data in Figure \ref{StokesEvolution.fig}.
In this framework, as a first, straightforward test for variability, we calculated the $\chi^{2}$ statistic for the mean $Q/U$ Stokes parameters in each time bin under the null hypothesis that they are all consistent with the time-integrated mean Stokes parameters ($\bar{Q}$ and $\bar{U}$). Since the Stokes parameters from each time bin (denoted as $Q_{i}$ and $U_{i}$, respectively, for time bin $i$) are normally distributed and independent, the quantity 
\begin{equation}
    \chi^{2} = \sum_{i} \frac{ \left( Q_{i} - \bar{Q} \right)^{2} + \left( U_{i} - \bar{U} \right)^{2}}{\sigma_{i}^{2} } 
\end{equation}
 \noindent
 under this null hypothesis follows a $\chi^{2}$ distribution with $2N_{\rm bins}-2=46$ degrees of freedom. We can therefore compare the observed value of $\chi^{2}$ with the expected values. 
 For Obs.\ 1, we find $\chi^{2} = 55$, 
 the probability of which arising from the null hypothesis of constant Stokes parameters is $P_{\rm null} \sim 0.17$, suggesting that these data are consistent with the constant polarization model. Conversely, for Obs.\ 2 and 3, we determine $\chi^{2}$ values of 78 and 121, respectively.
 The probability of acquiring $\chi^{2}$ exceeding this value by chance under the null hypothesis of constant polarization is $P_{\rm null} < 2 \times 10^{-3}$. This implies that, in these cases, the null hypothesis can be rejected with high confidence.\\
Using this same statistical framework, we are able to determine the best-fit rotation rate $\dot\Psi_{\rm x}$ for the model by assuming that both $\dot\Psi_{\rm x}$ and the polarization degree are constant (see above for the mathematical form of this model). 
When applying these calculations to the data of Obs.\ 1, we find no evidence of time-dependent rotation of $\Psi_{\rm x}$: the best-fit constant rotation rate is $-4 \pm 7$ \rotdeg). For Obs.\ 2 and 3, the best-fit rotation rates are $\dot\Psi_{\rm x} = 78\pm 8$ \rotdeg\ and $83\pm 6$ \rotdeg\, respectively. These values are consistent with the findings of the previously described methods using the same model, indicating that all three methods result in a consistent rate of rotation. The corresponding $\chi^{2}$ value for each of these fits is 53 for 45 $d.o.f$, with a corresponding value of $P_{\rm null} \sim 0.18$. We can therefore conclude that a constant rotation model provides a good fit to the data. Attempts to fit these data to a more complex elliptical rotation model (i.e., where the polarization degree varies with time) do not result in a statistically significant improvement in the overall value of $\chi^{2}$ given the reduced degrees of freedom.

\subsection*{Comparison with a stochastic model}
In order to verify further our explanation of the cause of the X-ray polarization angle rotation in \sou, we tested the hypothesis that the rotation occurred by chance as a result of a stochastic variation of $\Psi_{\rm x}$.
We considered two possibilities for the involvement of random processes. First, the variations in $\Psi_{\rm x}$ during the two IXPE observations of June corresponded to two independent rotations. Second, the two IXPE observations were part of one continuous rotation, of which we sampled only two segments. To determine the probability of the rotation(s) occurring by chance due to the random variations of $\Psi_{\rm x}$, we simulated the time
dependence of the polarization degree and angle that occur via random walks, and compared them to the IXPE results.
As described in detail in \cite{Kiehlmann2016, Kiehlmann2017}, the simulations are based on two parameters: the number of maximally polarized (i.e., each with uniform magnetic field) cells, $N_{\rm cell}$, and the number of cells that change per day, $N_{\rm Var}$. The cells each have the same intensity, hence the total intensity $I_{\rm total}=N_{\rm cell}I$. The initial Stokes parameters $Q$ and $U$ for each cell were drawn from Gaussian distributions. At every simulation time step, one cell was selected randomly, and its
$Q$, $U$ values were drawn anew, then the averages of $Q$, $U$, $\Pi$, and $\Psi$ over all cells were calculated anew. The number of simulated time steps depends on the observation duration and ($N_{\rm Var}$); e.g., for $N_{\rm Var}=100$ we simulated 100~data points per day. For each of the aforementioned possibilities, we performed a grid search in the $N_{\rm cell},~N_{\rm Var}$ parameter space to identify episodes of rotation of $\Psi$.
\\

\begin{figure}
\includegraphics[width=1\textwidth]{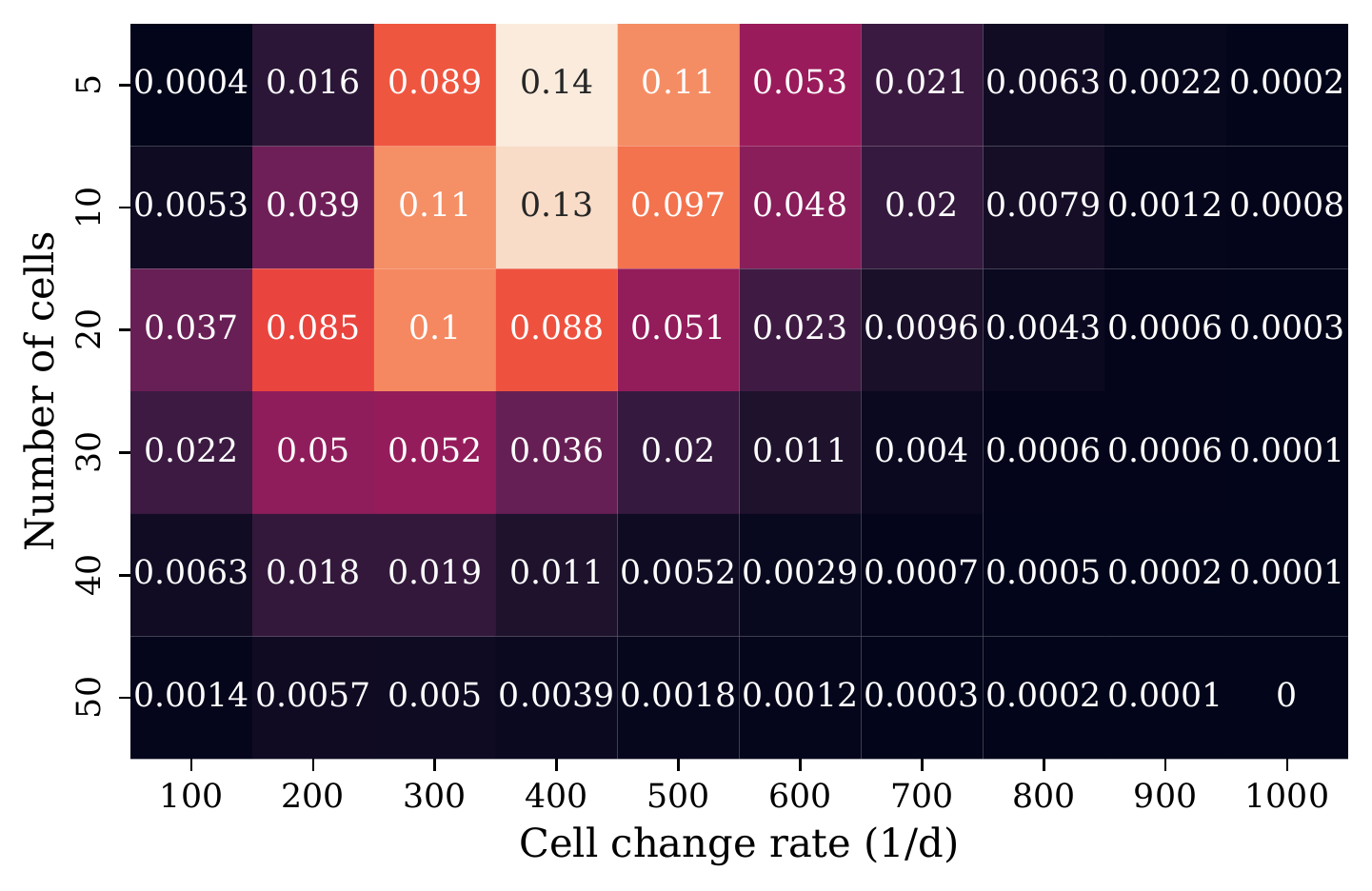}
 \caption{Random-walk parameter search for the second IXPE observation. The numbers in the squares indicate the fraction of successful trials.}
  \label{random_walk1}
\end{figure}

\begin{figure}
\includegraphics[width=1\textwidth]{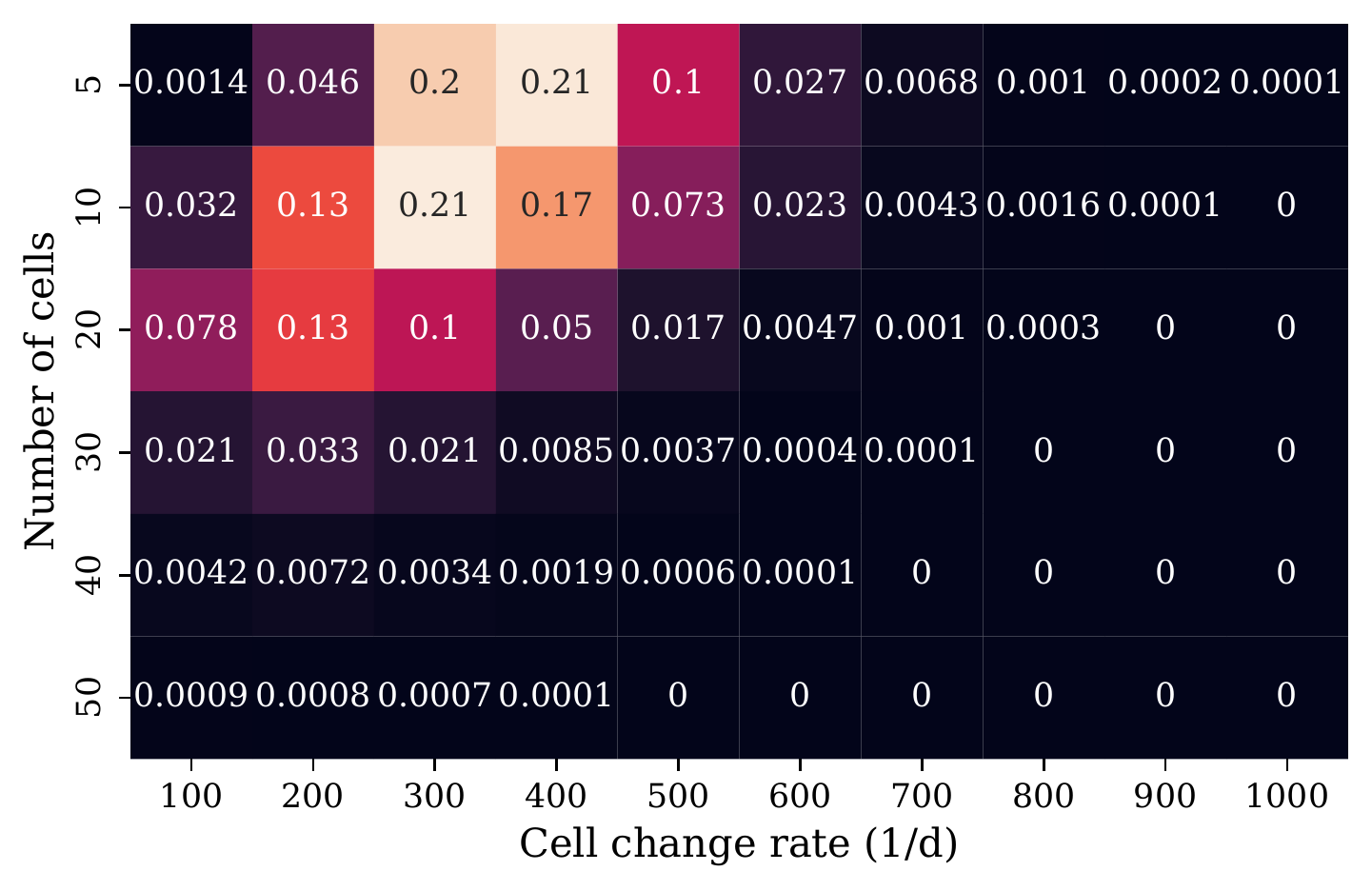}
 \caption{Random-walk parameter search for the third IXPE observation. The numbers in the squares indicate the fraction of successful trials.}
  \label{random_walk2}
\end{figure}

For the first case, we set the total length of the simulated polarization light curves to the length of each of the IXPE observations. We created 10,000 simulated light curves, binned to match one of the data binning schemes, and recorded the average $\Pi$ of the light curve and the longest continuous variation of $\Psi$ that could be interpreted as a rotation. We considered a trial successful if the average $\Pi$ and standard deviation of the simulated time dependence was within 10\% of the observed values, and if the variation in $\Psi$ was equal to or larger than that observed. Figures \ref{random_walk1} and \ref{random_walk2} show the results for the parameter search of each event. The binning of the data can have a significant effect on the number of successful trials. For that reason, we repeated the analysis for different numbers of bins from 3 to 18. For IXPE Obs.\ 2 (first rotation) we found the $p$ significance values to range from 0.0049 to 0.14 (13 bins), with a median of 0.057. For the following rotation (IXPE Obs.\ 3), we found $p=0.0013$ to 0.21 (17 bins), with a median of 0.032. The probability of two independent events occurring by pure chance in a row is then $p=0.029$. 
\\

\begin{figure}
\includegraphics[width=1\textwidth]{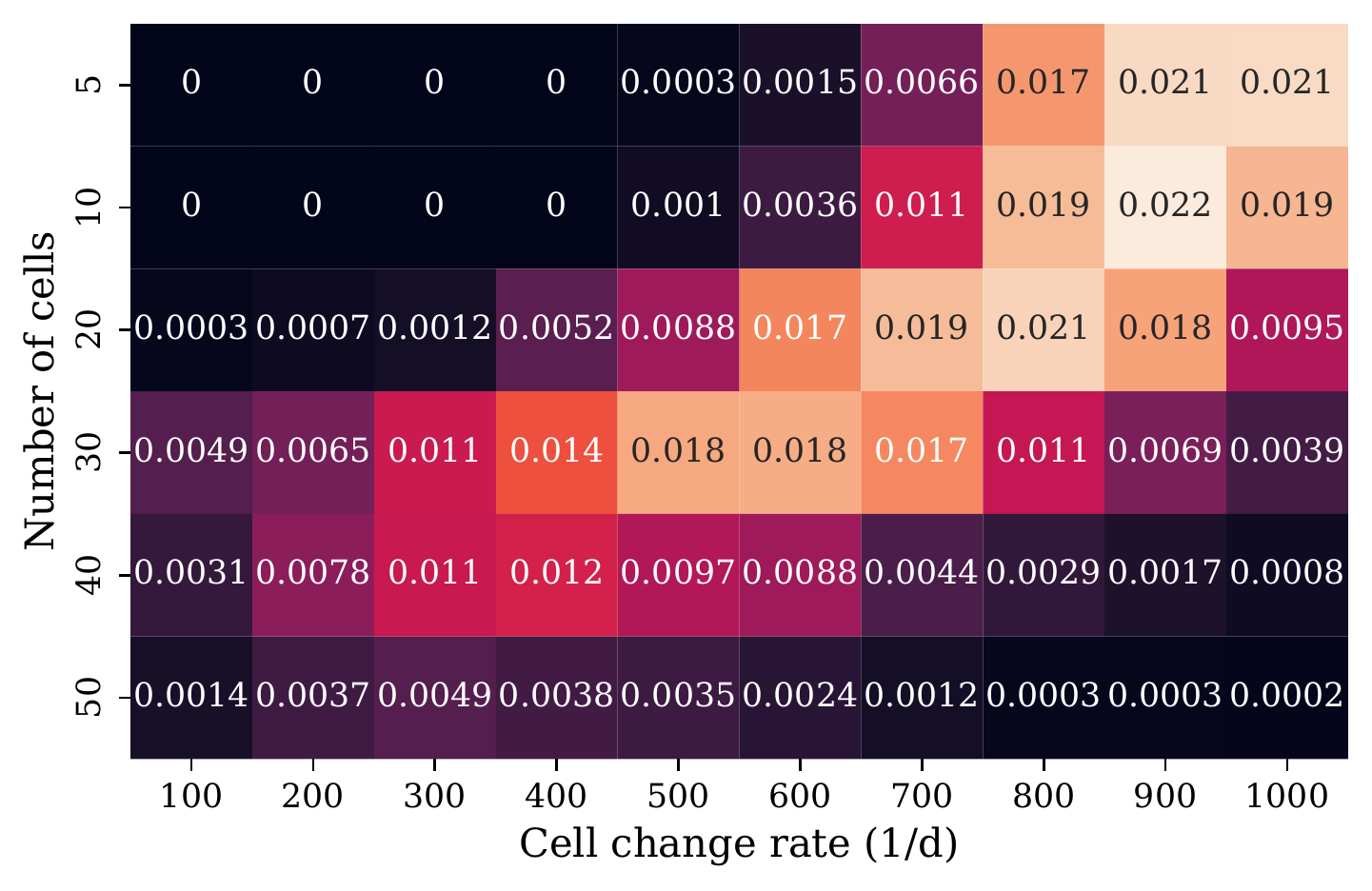}
 \caption{Random-walk parameter search assuming the second and third IXPE observations are part of the same rotation. The numbers in the squares indicate the fraction of successful trials.}
  \label{random_walk_both}
\end{figure}

For the second possibility, we set the length of the simulated light curves equal to the total duration of the observations from the beginning of IXPE Obs.\ 2 to the end of Obs.\ 3. To test the case of an individual rotation, the polarization-angle curve of Obs.\ 3 was shifted by 
$\pm 180^\circ$ as needed to align the two segments. In this case, we explored from 3 to 26 bins and additionally required that the rotation rate of the second segment lie within 10\% of that of the first segment. We repeated the above procedure and estimated the number of successful trials (Fig. \ref{random_walk_both}). We find $p=0.0006$ to 0.022 (23 bins) with a median of 0.0053. All of the above results indicate that it is unlikely that the observed rotation(s) were caused by a purely stochastic process. Rather, the event has a high probability of being related to non-stochastic physical processes in the jet.

\begin{figure}
\includegraphics[width=1\textwidth]{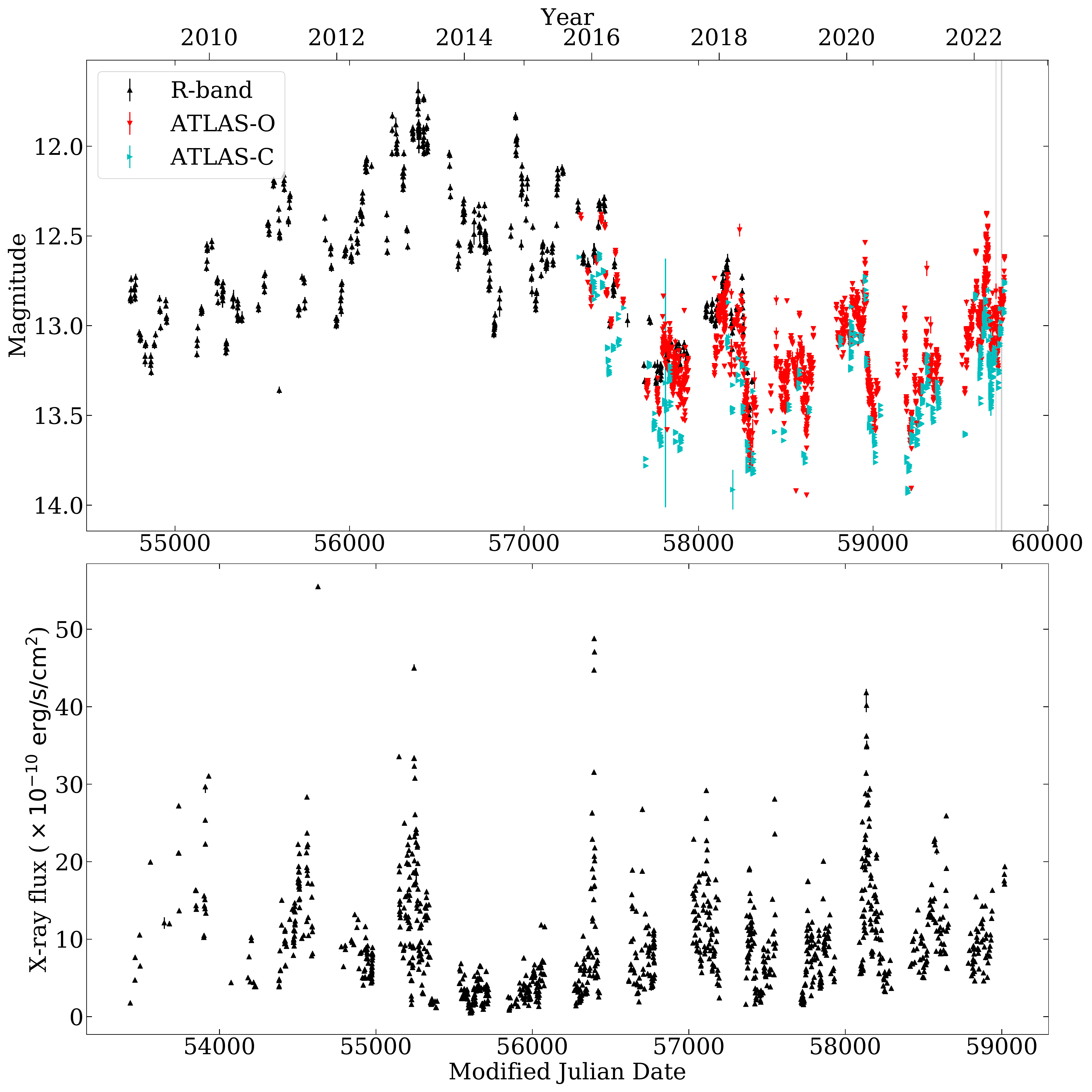}
 \caption{Archival optical brightness (black triangles: R-band, Steward observatory, red and cyan triangles: c-band and o-band, ATLAS observatory),
 and X-ray flux (lower panel, black triangles) light
curves for \sou. The grey shaded areas represent the duration
of the IXPE pointings. }
  \label{archive.fig}
\end{figure}

\subsection*{Activity state of \sou}
\label{state}
In Fig. \ref{archive.fig} we display the long term light curves of \sou\ in optical brightness (top panel) from the Steward observatory monitoring program \citep{Smith2009} and ATLAS \citep{Tonry2018,Heinze2018,Shingles2021}, as well as the X-ray flux (bottom panel) from the Swift XRT. The optical observations cover a time range from 2006 to 2022 and we highlight as vertical grey lines the time of the IXPE pointings.\\
The Swift-XRT observations are in the 0.3-10 keV band.  The minimum and maximum value of the X-ray flux were $0.5 \times 10^{-10}$ and $55 \times 10^{-10}$  \ergsc, respectively, with a median value of $8.5 \times 10^{-10}$. As a comparison, the flux measured by Swift at the time of the IXPE pointings (see Table \ref{xrtlog.tab}) ranged between $7.5 \times 10^{-10}$ and $14.1 \times 10^{-10}$ \ergsc. Thus, we conclude that \sou\ was at an average level of X-ray activity at the time of the IXPE observations. \\
We display the optical brightness in the top panel of Fig. \ref{archive.fig}. 
The Steward observatory observations are in the R-band, while for ATLAS we show both c-band and o-band observations. While the Steward and ATLAS observations are not directly comparable without applying conversion factors between optical bands, it is evident from the overlap of the two surveys (2016-2018) that any difference is rather small compared to the overall brightness variations of \sou. The ATLAS observations in 2022 are $\sim$0.5mag lower than the average of the Steward observations (2011-2016) and $\sim$1-1.5mag lower than the flares seen in the Steward data. Therefore, we can also conclude that the optical brightness of \sou\ at the time of the IXPE observations was at an average state.

\end{document}